\documentclass{article}
\usepackage{graphicx}
\usepackage{natbib}

\usepackage{amsmath}
\usepackage{amssymb}

\begin{document}

\title{Massive identification of asteroids in three-body resonances}

\author{E. A. Smirnov and I. I. Shevchenko\/\thanks{E-mail: iis@gao.spb.ru} \\
Pulkovo Observatory of the Russian Academy of Sciences \\
Pulkovskoje ave.~65, St.\,Petersburg 196140, Russia}
\date{}

\maketitle

\begin{abstract}
An essential role in the asteroidal dynamics is played by the mean
motion resonances. Two-body planet-asteroid resonances are widely
known, due to the Kirkwood gaps. Besides, so-called three-body
mean motion resonances exist, in which an asteroid and two planets
participate. Identification of asteroids in three-body (namely,
Jupiter-Saturn-asteroid) resonances was initially accomplished by
D.\,Nesvorn\'y and A.\,Morbidelli (1998), who, by means of visual
analysis of the time behaviour of resonant arguments, found 255
asteroids to reside in such resonances. We develop specialized
algorithms and software for massive automatic identification of
asteroids in the three-body, as well as two-body, resonances of
arbitrary order, by means of automatic analysis of the time
behaviour of resonant arguments. In the computation of orbits, all
essential perturbations are taken into account. We integrate the
asteroidal orbits on the time interval of 100000~yr and identify
main-belt asteroids in the three-body Jupiter-Saturn-asteroid
resonances up to the 6th order inclusive, and in the two-body
Jupiter-asteroid resonances up to the 9th order inclusive, in the
set of $\sim 250000$ objects from the ``Asteroids -- Dynamic
Site'' (AstDyS) database. The percentages of resonant objects,
including extrapolations for higher-order resonances, are
determined. In particular, the observed fraction of pure-resonant
asteroids (those exhibiting resonant libration on the whole
interval of integration) in the three-body resonances up to the
6th order inclusive is $\approx 0.9\%$ of the whole set; and,
using a higher-order extrapolation, the actual total fraction of
pure-resonant asteroids in the three-body resonances of all orders
is estimated as $\approx 1.1\%$ of the whole set.
\end{abstract}



\section{Introduction}

A substantial role of resonances in the dynamics of asteroids
became evident with the discovery of resonant ``gaps'' in the
asteroid belt by D.\,Kirkwood in 1867. The deepest minima in the
distribution of asteroids in the semimajor axes of their orbits
correspond to the mean motion resonances 2/1, 3/1, 4/1, 5/2, and
7/3 with Jupiter. Mean motion resonance represents a
commensurability between the mean frequencies of the orbital
motions of an asteroid and a planet. Apart from the mean motion
resonances, so-called secular resonances \citep{MD99,M02},
representing commensurabilities between the precession rates of
the orbits of an asteroid and a planet, are important in forming
the dynamical structure of the asteroid belt.

There are two important classes of the mean motion resonances:
apart from the usual (two-body) mean motion resonances of an
asteroid and a planet, an appreciable role in the asteroidal
dynamics is played by so-called three-body mean motion resonances
\citep{MHP98,NM98,NM99,M02}. In the latter case, the resonance
represents a commensurability between the mean frequencies of the
orbital motions of an asteroid and two planets (e.g., Jupiter and
Saturn):

\begin{equation}
  \label{dr}
  m_\mathrm{J}\dot{\lambda_\mathrm{J}}+m_\mathrm{S}\dot{\lambda_\mathrm{S}}+
  m\dot{\lambda} \approx 0,
\end{equation}

\noindent where $\dot{\lambda_\mathrm{J}}$,
$\dot{\lambda_\mathrm{S}}$, $\dot{\lambda}$ are the time
derivatives of the mean longitudes of Jupiter, Saturn, and
asteroid, respectively, and $m_\mathrm{J}$, $m_\mathrm{S}$, $m$
are integers.

In view of the ``overdensity'' of the three-body resonances in the
phase space of the asteroidal motion, \cite{NM98} asserted that
``the three-body mean motion resonances seem to be the main actors
structuring the dynamics in the main asteroid belt''.

Chaotic behaviour, which is often present in the dynamics of
celestial bodies, is usually due to interaction of resonances (as
in any Hamiltonian system, see \citealt{C79}), but not always it
is known which are the interacting resonances that give rise to
chaos. It is especially difficult to identify three-body
resonances. How to distinguish between resonant and non-resonant
motions? To solve this problem, a ``resonant argument''
(synonymously ``resonant phase'' or ``critical argument'') is
introduced. It is a linear combination of some angular variables
of a system under consideration; in the planar asteroidal problem
it is given by

\begin{equation}
  \label{radinition}
  \sigma_{p_\mathrm{J}, p_\mathrm{S}, p}=m_\mathrm{J}\lambda_\mathrm{J}+
  m_\mathrm{S}\lambda_\mathrm{S}+m\lambda+p_\mathrm{J}\varpi_\mathrm{J}+
  p_\mathrm{S}\varpi_\mathrm{S}+p\varpi,
\end{equation}

\noindent where $\lambda_\mathrm{J}$, $\lambda_\mathrm{S}$,
$\lambda$, $\varpi_\mathrm{J}$, $\varpi_\mathrm{S}$, $\varpi$ are
the mean longitudes and longitudes of perihelia of Jupiter,
Saturn, and an asteroid, respectively, and $m_\mathrm{J}$,
$m_\mathrm{S}$, $m$, $p_\mathrm{J}$, $p_\mathrm{S}$, $p$ are
integers satisfying the D'Alembert rule \citep{M02}:

\begin{equation}
  \label{dalambert_rule}
  m_\mathrm{J} + m_\mathrm{S} + m + p_\mathrm{J} + p_\mathrm{S} + p = 0 .
\end{equation}

If resonant argument~(\ref{radinition}) librates (similarly to
librations of a pendulum), the system is in resonance; if it
circulates, the system is out of resonance. The motion of the
system at the border between librations and rotations corresponds
to the {\it separatrix}. Thus the pendulum dynamics provides a
graphical model of resonance. In a certain sense this model of
resonance is ``universal'' \citep{C79}. In particular, the motion
in three-body resonances can be described in the perturbed
pendulum model \citep{MHP98, NM98, NM99, S07}.

An important parameter of a mean motion resonance is its {\it
order} $q$, equal to the absolute value of the algebraic sum of
the coefficients at the mean longitudes in the resonant argument:

\begin{equation}
q = |m_\mathrm{J} + m_\mathrm{S} + m| .
\end{equation}

\noindent The resonant order $q$ is important, because it is the
power in which the eccentricity is raised in the coefficient of
the leading resonant term in the expansion of the perturbing
function~\citep{NM98}. The corresponding subresonance width
(characterizing also its ``strength'') is proportional to the
square root of this coefficient. Thus the value of $q$ determines
this important property of the leading subresonance.

In the case of two-body resonances, the role of the resonant order
$q$ (defined below in Section~\ref{sec_2br}) is analogous: the
coefficient of the leading resonant term is proportional to $e^q$
\citep{NM98}, where $e$ is the asteroidal eccentricity.

However note that, when there is no strong overlapping of
subresonances, the resonant order $q$ is not related to the width
of the whole resonant multiplet, because the separation of
subresonances depends solely on the secular precession rates of
the pericentres \citep{NM99}; thus the degree of overlap (and
hence, chaos) in the multiplets is expected to asymptotically
decrease with the resonant order \citep{NM99,MG97}.

One may expect that, generally, broader the leading subresonance
of a mean motion resonance, greater is the number of objects
residing in this mean motion resonance. However, no strict
correlation exists, due to a competition of various dynamical and
physical processes, populating or depopulating the resonances. We
shall discuss this further in more detail.

In our procedure of resonance identification, described in detail
below, we limit the set of possible combinations of the integers
$m_\mathrm{J}$, $m_\mathrm{S}$, $m$ by adopting the following
conditions:

\begin{equation}
  \label{iac}
  q \leq q_\mathrm{max} ,
\end{equation}

\begin{equation}
  \label{ic1}
  |m_\mathrm{J}|, \ |m_\mathrm{S}|, \ |m| \leq M_\mathrm{max},
\end{equation}

\noindent where $q_\mathrm{max} = 6$ and $M_\mathrm{max} = 8$.

\cite{NM98} used condition~(\ref{iac}) (presumably with
$q_\mathrm{max} = 10$, as follows from data in table~3 in
\citealt{NM98}). In \citep{NM99}, instead of (\ref{ic1}), the
following truncation condition was used:

\begin{equation}
  \label{inc}
  |m_\mathrm{J}|+|m_\mathrm{S}|+|m| \leq Q_\mathrm{max}
\end{equation}

\noindent (see eqs.~(29) and (30) and comments on them in
\citealt{NM99}).

We identify the three-body resonances in the current motion of
asteroids with known orbital elements. The limitations of our
study are as follows: solely the asteroids in the main belt are
considered (i.e., the semimajor axes are in the range from 2 to
4~AU); solely the three-body resonances with Jupiter and Saturn are
taken into account; the resonances are considered in the planar
problem, i.e., the longitudes of nodes in the expression for the
resonant argument are ignored; the maximum considered order
$q_\mathrm{max}$ of the three-body resonances is set equal to 6.

Our project is intended for the resonance analysis of the orbital
data presented at the ``Asteroids -- Dynamic Site'' (AstDyS)
maintained by A.\,Milani, Z.\,Kne\v{z}evi\'c and their coworkers
\citep{AstDyS}. We take the orbital data for the
analysis from this database. Thus the total set under analysis
contains $\approx 250000$ objects.

The basic purpose of our work is to identify the current
three-body resonances that all the asteroids from the given set
are currently involved in. More specifically, each object from the
set should be put in correspondence to a three-body resonance (or
none, if there is no resonance). The first attempt of massive
identification of asteroids in three-body resonances was made by
\cite{NM98}: 255 objects were identified to be in three-body
resonances. The libration/circulation of the resonant argument for
asteroids suspected to reside in the resonances was analyzed
visually. In our case the data set is much greater, and therefore
the procedure ought to be completely automatic.
Besides, here we apply a unified bound on the order. This allows
one to construct a homogeneous identification list for a further
statistical analysis.

To form a general statistical view of the resonant structure of
the main belt, we also accomplish a massive identification of
asteroids in two-body resonances with Jupiter, and compare the
abundances of asteroids in three-body and two-body resonances.

\section{The identification matrix}
\label{scim}

As a first stage of the identification process we build an
``identification matrix''. It consists of two main columns. The
first one contains designations of resonances, and the second one
contains the corresponding resonant values of the semimajor axis.

The designations of resonances are given in the notation
$m_\mathrm{J} m_\mathrm{S} m (q)$, where $m_\mathrm{J}$,
$m_\mathrm{S}$, $m$ are the integer coefficients in the resonant
argument~(\ref{radinition}), and $q$ is the resonant order, as
defined above. The values of $m_\mathrm{J}$, $m_\mathrm{S}$, $m$
are given with their signs. Thus, examples of this notation look
like as follows: 5-2-2(1), 2+2-1(3).

We construct a set of the resonant arguments for all possible
three-body resonances up to a fixed order $q_\mathrm{max}$ in the
following way.

We fix the maximum absolute value $M_\mathrm{max}$ of each integer
$m_\mathrm{J}$, $m_\mathrm{S}$, $m$ to be equal to $q_\mathrm{max}
+ 2$, where $q_\mathrm{max}$ is the maximum resonant order. It is
assumed that

\begin{equation}
m_\mathrm{J}>0, \quad \gcd{(m_\mathrm{J}, m_\mathrm{S}, m)}=1,
\end{equation}

\noindent where ``gcd'' stays for the greatest common divisor. It
is set to be equal to 1 to avoid the higher order harmonics with
greater {\it multiplicity}. The multiplicity is defined as equal
to $\gcd{(m_\mathrm{J}, m_\mathrm{S}, m)}$. The harmonics with
multiplicity greater than 1 are not discernible in our
identification procedure, because their arguments librate
simultaneously, though with different amplitudes. (Consider, e.g.,
such resonances as $4-2-2$ and $8-4-4$. The second one, which has
multiplicity equal to 2, in our procedure is set to be equivalent
to the first one.)

Then we search through all possible combinations of
$m_\mathrm{J}$, $m_\mathrm{S}$, $m$ and identify those satisfying
the D'Alembert rule~(\ref{dalambert_rule}) and our technical
restrictions (\ref{iac}) and (\ref{ic1}).

Let us demonstrate how the resonant value of the semimajor axis is
calculated. According to the definition of the three-body
resonance~(\ref{dr}), the time derivative
$\dot{\sigma}_{p_\mathrm{J}, p_\mathrm{S}, p}$ should be equal to
zero. Let us, following \cite{MHP98}, assume that
$\dot{\varpi}\approx 0$, in the first approximation. Then, for the
resonant value of mean motion, one has

\begin{equation}
  \label{mmfm}
  n_\mathrm{res} = - \frac{1}{m}(m_\mathrm{J}\dot{\lambda_\mathrm{J}} +
  m_\mathrm{S}\dot{\lambda_\mathrm{S}} + p_\mathrm{J}\dot{\varpi}_\mathrm{J} +
  p_\mathrm{S}\dot{\varpi}_\mathrm{S}) .
\end{equation}

\noindent Using Kepler's third law, one obtains for the resonant
semimajor axis

\begin{equation}
\label{afmm}
  a_\mathrm{res} = \left(\frac{k}{n_\mathrm{res}}\right)^{2/3} ,
\end{equation}

\noindent where $n_\mathrm{res}$ is given by formula~(\ref{mmfm}),
and $k$ is the Gauss constant.

\cite{HM96} and \cite{MHP98} obtained an approximate formula for
the precession rate of an asteroid's orbit:

\begin{equation}
\label{lfmu}
  \dot{\varpi}\approx \frac{\mu}{2\pi}
  \left( \frac{a}{a_\mathrm{J}} \right)^{1/2}
  \varepsilon^2n_\mathrm{J},
\end{equation}

\noindent where $\mu \approx 1/1047$ is the mass of Jupiter in
units of the mass of the Sun, $n_\mathrm{J}$ and $a_\mathrm{J}$
are Jupiter's mean motion and the semimajor axis of Jupiter's
orbit, respectively, and

\begin{equation}
  \varepsilon = \frac{a_\mathrm{J}-a}{a_\mathrm{J}} .
\end{equation}

\noindent For $a$, we substitute here the resonant value of the
semimajor axis as given by Eq.~(\ref{afmm}). Thus the value of
$\dot{\varpi}$ is calculated. Iterating, one obtains an adequate
value of $a_\mathrm{res}$. The second primary column of the
identification matrix is filled with the resonant values of the
semimajor axes, calculated as described, with the accuracy of no
less than $10^{-3}$~AU. This accuracy far exceeds the necessary
one, because we check the asteroids for belonging to a given
resonance in a far greater neighborhood ($\pm 10^{-2}$~AU) of the
computed resonant value of $a_\mathrm{res}$.

\cite{NM99} calculated $a_\mathrm{res}$ of the leading
subresonances (i.e., of the multiplet components $\sigma_{0, 0,
-m_\mathrm{J} -m_\mathrm{S} -m}$) of 19 three-body resonances with
the accuracy of $10^{-4}$~AU: they equated the time derivative of
Eq.~(\ref{radinition}) to zero, and, using the values of $\dot
\lambda_\mathrm{J}$ and $\dot \lambda_\mathrm{S}$ as given by
\cite{B90} and $\varpi$ as found using the code by \cite{MK94},
calculated $\dot \lambda$. All values of $a_\mathrm{res}$ given in
\cite[table~1]{NM99} agree quite closely with the corresponding
values of $a_\mathrm{res}$ calculated here iteratively, as
described above, for our matrix. The agreement is illustrated in
Table~\ref{tims}, where an extract from the identification matrix
is presented.

\begin{table}[ht]
\caption{An extract from the identification matrix}
\medskip
\centering
\begin{tabular}{rrrrcc}
\hline
$m_\mathrm{J}$ & $m_\mathrm{S}$ & $m$ & $q$ & $a_\mathrm{res}$ (AU) & $a_\mathrm{res}$ (AU) \\
            &             &     &     &        (this study) & \protect\citep{NM99} \\
\hline
2 & 3 & -1 & 4 & 2.3912 & --- \\
4 & -2 & -1 & 1 & 2.3978 & 2.3977 \\
2 & 2 & -1 & 3 & 2.6148 & 2.6155 \\
4 & -3 & -1 & 0 & 2.6232 & 2.6229 \\
1 & 4 & -1 & 4 & 2.7432 & --- \\
3 & -1 & -1 & 1 & 2.7527 & 2.7525 \\
1 & 3 & -1 & 4 & 3.0673 & --- \\
3 & -2 & -1 & 0 & 3.0798 & 3.0790 \\
5 & -7 & -1 & 3 & 3.0925 & --- \\
5 & -2 & -2 & 1 & 3.1744 & 3.1751 \\
\hline \label{tims}
\end{tabular}
\end{table}

\section{Dynamical identification}

We use the following procedure of dynamical identification.

First of all, each asteroid's orbit from the adopted set of 249567
objects is computed for $10^5$~yr. The perturbations from all
planets (from Mercury to Neptune) and Pluto are taken into
account. The hybrid integrator of {\it mercury6} package
\citep{Mercury6} is used. In some cases the {\it orbit9}
integrator \citep{OrbFit} is used as well, to verify the results.
The computed trajectories are kept in files for further usage. The
trajectories are output with the time step of 1~yr.

After integration is over, the objects are taken from
the adopted set, and the mean value of the semimajor axis is
computed for each object. Using this value, a set of preliminary
resonant arguments $\sigma_\mathrm{res}$ is found in the
identification matrix. Each argument $\sigma_\mathrm{res}$ is then
analyzed on the presence of libration/circulation, using the
computed trajectory of the object.

\begin{figure}[!ht]
  \includegraphics[width=1.0\textwidth]{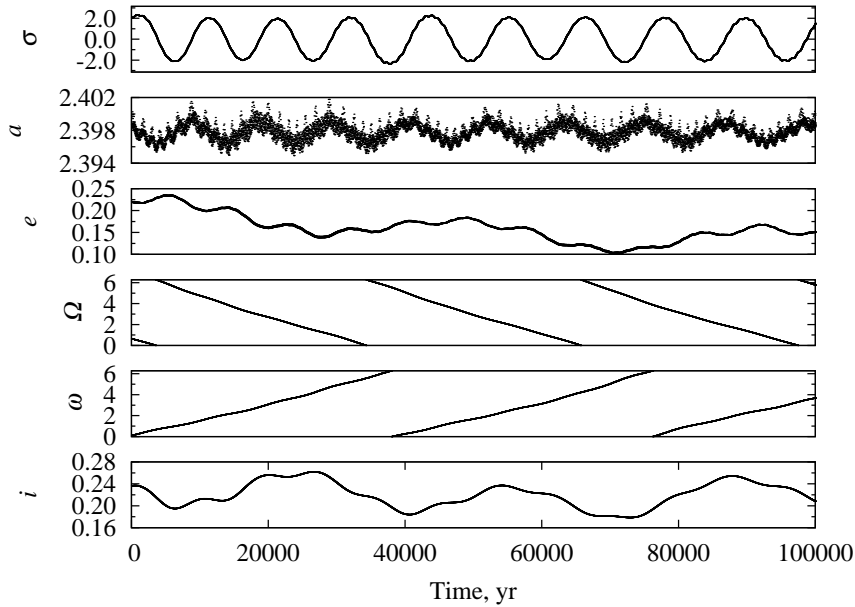}
  \caption{The orbital elements and resonant argument
  of 463~Lola. Resonance $4 -2 -1$.}
  \label{f463e}
\end{figure}

We distinguish two types of resonant libration: pure and
transient. By definition, the libration is pure, if it lasts
during the whole time interval of integration, i.e., $10^5$~yr. An
example of such libration is given in Fig.~\ref{f463e}, where the
time behaviour of the resonant argument, alongside with the
orbital elements, is demonstrated for asteroid 463~Lola, resonance
$4 -2 -1$.

\begin{figure}[!ht]
    \includegraphics[width=1.01\textwidth]{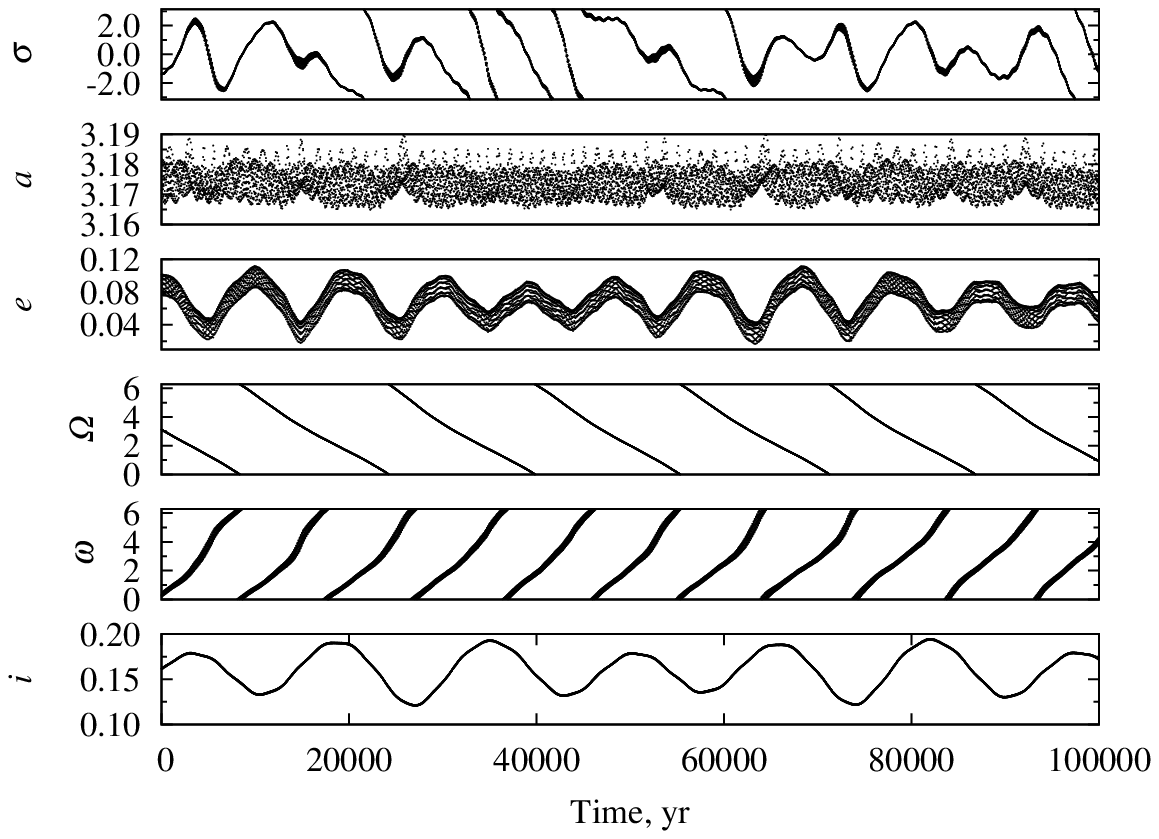}
  \caption{The orbital elements and resonant argument
  of 490~Veritas. Resonance $5 -2 -2$.}
  \label{f490e}
\end{figure}

The libration is defined as transient, if circulation appears at
any time during this interval. An example is given in
Fig.~\ref{f490e}; this is the case of asteroid 490~Veritas,
resonance $5 -2 -2$. In Fig.~\ref{f490e}, it may seem unusual that
Veritas exhibits circulation of the resonant argument while the
semimajor axis remains almost constant (especially in comparison
with Fig.~\ref{f1915twb}, which is considered below and which
shows the orbital elements and resonant argument of 1915
Quetz\'alcoatl, residing in the two-body transient resonance 3/1).
This difference is explained by a large difference in the widths
of the resonances: e.g., at $e=0.1$ the width of resonance 3/1 is
$\sim 5$ times greater than that of resonance $5 -2 -2$ (see
fig.~1 in \citealt{MN99}). Moreover, 1915 Quetz\'alcoatl has a
very large eccentricity ($\sim 0.6$--$0.8$). At $e=0.1$, the
half-width of the $5 -2 -2$ resonance is $\approx 0.002$~AU (see
table~1 in \citealt{NM99} and fig.~1 in \citealt{MN99}). For
Veritas, the eccentricity is $\sim 0.06$ (see Fig.~\ref{f490e}),
hence the half-width is even smaller. This makes the shift in $a$,
when $\sigma$ is circulating, almost imperceptible, especially in
the digitally unfiltered $a$. When $a$ is filtered, such shifts
look more obvious; see fig.~2 in \citep{KTV02} and fig.~3 in
\citep{TKV07}. Note that, when the circulations are short-term,
the shifts in $a$ can be imperceptible even in the digitally
filtered element: see fig.~1 in \citep{NM98}, where the time
behaviour of Veritas in the digitally filtered elements is shown.

To distinguish between transient-resonant and non-resonant
behaviours we introduce a technical parameter: the {\it resonance
minimum time}, which we set to be equal to 20000~yr. Roughly
speaking, this is the minimum time interval of resonant
librations, for an asteroid to be regarded as resonant. An exact
definition is given below.

\cite{NM98} identified resonant behaviour visually and used more
subjective criteria, according to which the asteroid is resonant
if ``(1)~the corresponding resonant angle shows evident librations
during the integration time span or (2) the resonant angle
circulates with a period longer than several thousand years''
\citep{NM98}. In fact the second criterion threshold is analogous
to our resonance minimum time, because our automatic procedure
measures the time intervals of libration as such when no
circulation is present, as described in detail below. Our
criterion threshold, being larger, is more restrictive; however,
in both cases the threshold is much greater than the timescales of
circulation far from resonance; the latter timescales are of the
order of asteroidal orbital periods.

The complete procedure of identification is as following. When
computing each trajectory, the resonant argument value for the
guiding subresonance is calculated at each step in time, and these
values are written in a file, in function of time. After the
trajectory computation is over, the time behaviour of the resonant
argument is analyzed. If a current value of the resonant argument
is different from its previous value by more than $2 \pi$, this
means that there is a break, which takes place if there is either
circulation or apocentric libration. To distinguish between these
two cases, the same procedure is repeated, but with an artificial
shift of the resonant argument. This shift is equal to $\pi$.
Then, if there is apocentric libration, it turns into pericentric
one, and there are no breaks. At the beginning of the trajectory
analysis, two variables are initiated: the first one is the
duration of current libration, and the second one is the total
time of libration. If circulation whenever starts, the first
variable value is added to the second one and the first variable
is reset to zero. At the end of the analysis, it is checked
whether the second variable (total time of libration) is equal to
the full time of computation ($10^5$~yr). If yes, then the
asteroid is regarded to be in {\it pure resonance}. If no, but the
second variable value exceeds the {\it resonance minimum time},
then the asteroid is in {\it transient resonance}. If the second
variable value is less than the {\it resonance minimum time}, then
the asteroid is regarded to be non-resonant. The adopted
pure/transient division of resonances is thus mostly technical,
because it depends on the chosen time interval of computation.

We use solely the direct method of identification, i.e., the
resonant argument is checked on the subject of libration. Any
secondary and/or auxiliary criteria, such as the semimajor axis
behaviour, the location of objects in the proper ``semimajor axis
-- eccentricity'' plane, the Lyapunov exponents are left for
future explorations, --- with a partial exception in the latter
case, see Section~\ref{Lestat}.

In our identification procedure, it is formally possible that an
asteroid might be identified as belonging to two or even more
resonances, because all resonances in a rather broad neighbourhood
in semimajor  axis ($\pm 10^{-2}$~AU, as mentioned above) in the
identification matrix are scanned on the subject of libration. If
two three-body resonances are close enough in semimajor axis, and
especially if they also overlap with a strong two-body resonance,
an asteroid may intermittently diffuse from one to another
resonance; a small number of such objects have been identified, as
belonging to two resonances. In our lists, each object of this
kind has been finally attributed to the resonance where it stayed
for a longer time, to avoid complication of statistics. Analysis
of such objects (we call them ``rogue-resonant'' asteroids) will
be given elsewhere.

Besides, since the data used is unfiltered, some
misidentifications may take place when the libration amplitude is
high. To estimate the formal statistical error of our procedure,
we randomly chose 300 asteroids identified as pure-resonant and
visually checked the libration of their resonant arguments. It
turned out that only 4 objects ($1.3\%$ of the set) were
misidentified. Thus the statistical error of the procedure is
$\approx 1\%$.

All identified pure three-body-resonant asteroids, grouped
according to association to a given resonance, are listed in the
Appendix~A. The top ten resonances, that are most ``populated'',
are listed in Table~\ref{trt10}. The last column of
Table~\ref{trt10} contains analytical estimates of the resonance
width (at the asteroidal eccentricity $e=0.1$), according to
\cite[table~1]{NM99}. One can see that, generally, broader the
resonance, greater is the number of objects residing in it.
However, this tendency is not strict. The reason is that the
dynamics here is strongly interrelated with physics: e.g., a
collisional disintegration of an asteroid can strongly increase
abundance of objects in a particular resonance, thus disturbing
the expected correlation.

For each resonance, statistics on the asteroids in pure and
transient librations have been calculated. The statistical results
are summarized in Table~\ref{trs}. The fraction of asteroids in
three-body resonances (transient plus pure) turns out to be
$\approx 4.4\%$ of the whole set. This is rather close to the
value $4.6\%$ ($\approx 1500$ in the set of $\approx 32400$) found
by \cite{NM98} to serve as a lower bound for the relative number
of resonant asteroids. As follows from Table~\ref{trs}, the
fraction of asteroids in pure three-body resonances turns out to
be $\approx 0.94\%$ of the whole set.

The third column of Table~\ref{trs} contains a prediction for
the numbers of asteroids residing in resonances of all orders;
this subject is discussed in the next Section.

\begin{figure}
  \centering
  \includegraphics[width=0.8\textwidth]{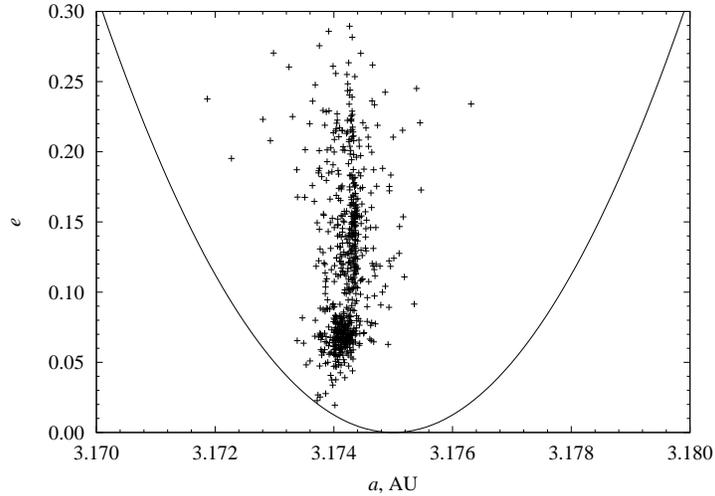}
  \caption{The asteroids identified to be in resonance $5 -2 -2$:
  location in the ``$a$--$e$'' plane.
  Solid curve: the separatrix of the leading subresonance, as calculated in
  \protect\citep{NM99}.}
  \label{s522}
\end{figure}

Let us consider the location of three-body-resonant asteroids in
the ``semimajor axis --- eccentricity'' plane. For each asteroid
in a given resonance we calculate the average values (over the
whole time interval of integration) of the semimajor axis and
eccentricity and plot these values in the ``$a$--$e$'' plane. We
have accomplished this procedure for each resonance with known
structure of the resonant multiplet, as calculated by \cite{NM99},
so that the separatrix of the leading subresonance could be drawn.
An example of such a plot is given in Fig.~\ref{s522}. It is clear
that the percentage of ``outliers'' (objects out of the separatrix
cell) is zero. This confirms the good accuracy of the accomplished
identification procedure in the case of this particular resonance.
A detailed study of the ``$a$--$e$'' plots will be given
elsewhere.

We have cross-checked the lists of resonant objects identified in
our study and in \citep{NM98}. It turns out that the number of
resonant objects listed by \cite{NM98} but not identified as
resonant in our study does not exceed $1\%$ of the list of
\cite{NM98}. This confirms that the differences in the
methodologies of identification in the two studies do not play any
significant role in what concerns the reliability of results.

\begin{table}[!ht]
\caption{The top ten most populated three-body resonances
Sun--Jupiter--asteroid}
\medskip \centering
\begin{tabular}{rrrcccc}
  \hline
  $m_\mathrm{J}$ & $m_\mathrm{S}$ & $m$ & $a_\mathrm{res}$ & Trans. + pure,\ & Pure, number & Res.\ width (AU) at $e=0.1$ \\
              &             &     &          (AU) & number of objects  & of objects & \protect\citep{NM99} \\
  \hline
  5 & -2 & -2 & 3.1744 & 699 & 182 & 0.0056\\
  4 & -2 & -1 & 2.3978 & 688 & 595 & 0.0024\\
  3 & -2 & -1 & 3.0798 & 621 & 134 & 0.0045\\
  3 & -1 & -1 & 2.7527 & 540 & 203 & 0.0019\\
  2 & 2  & -1 & 2.6148 & 470 &  34 & 0.00015\\
  4 & -3 & -1 & 2.6232 & 455 &  90 & 0.0009\\
  2 & 3  & -1 & 2.3912 & 343 &  56 & -- \\
  5 & -7 & -1 & 3.0925 & 314 &  48 & -- \\
  1 & 3  & -1 & 3.0673 & 300 &  52 & -- \\
  1 & 4  & -1 & 2.7432 & 284 &  47 & -- \\
  \hline
\label{trt10}
\end{tabular}
\end{table}

\begin{table}[!t]
\label{trs}
\caption{Asteroids in three-body resonances, statistics}
\medskip
\begin{center}
\begin{tabular}{lrrr}
  \hline
  & NM98$^*$ & This study, & This study, prediction \\
  & & $q_\mathrm{max} = 6$ & for $q_\mathrm{max} = \infty$ \\
  \hline
  The whole set of objects         & 5400  & 249567 & 249567\\
  Objects with integrated orbits & 836 & 249567 & 249567 \\
  Transient$+$pure-resonant objects & 255 & 11039 & --- \\
  The same, fraction of the studied set & 4.7\%  & 4.4\% & --- \\
  Pure-resonant objects          & ---  & 2338 & 2854 \\
  The same, fraction of the studied set & ---  & 0.94\% & $1.1$\% \\
  \hline
\end{tabular}
\end{center}
{\small $^*$\protect\cite{NM98}.}
\end{table}

\section{Expected abundances of asteroids in high-order three-body resonances}

The obtained list of resonant asteroids is obviously not complete,
due to the limitations of identification criteria. First of all,
the resonant order $q \leq 6$. Presumably, there is a lot of
objects in resonances of higher orders. To take account of them,
let us analyze the asteroid distribution in the resonant order
$q$. The necessary data as derived from the results of the
identification procedure are presented in Table~\ref{trod}.

\begin{figure}[!ht]
\includegraphics[width=1.0\textwidth]{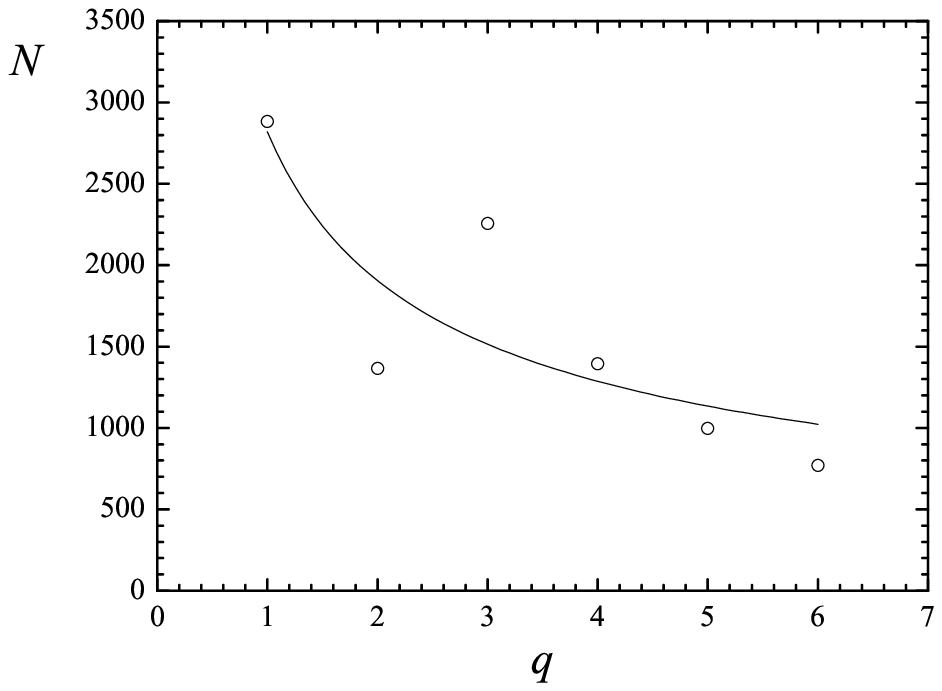}
  \caption{Circles: the distribution of all resonant (transient-resonant plus
  pure-resonant) asteroids in resonant order $q \geq 1$.
  Solid curve: the power-law fitting.}
  \label{froa}
\end{figure}

\begin{figure}[!ht]
\includegraphics[width=1.0\textwidth]{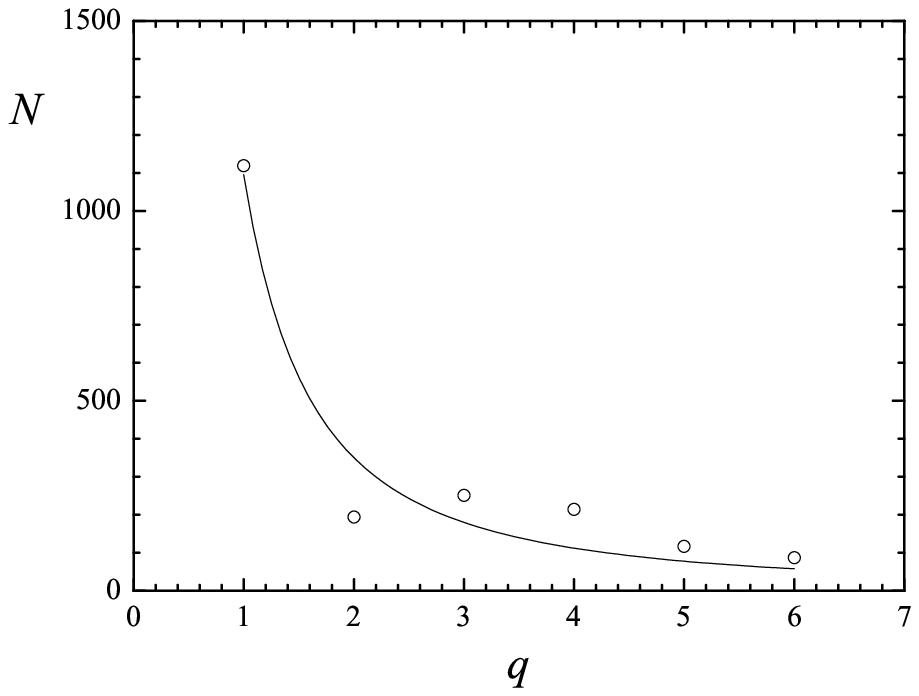}
  \caption{Circles: the distribution of pure-resonant asteroids
  in resonant order $q \geq 1$.
  Solid curve: the power-law fitting.}
  \label{fproan}
\end{figure}

The constructed differential distributions (histograms) in $q$ are
shown in Fig.~\ref{froa} for the case of transient plus pure
resonances, and in Fig.~\ref{fproan} for the case of pure
resonances.

\begin{table}[!t]
\caption{Abundances of asteroids in three-body resonances in
function of $q$}
\medskip \centering{
  \begin{tabular}{lrrrrrrr}
  \hline
  $q$ & 0 & 1 & 2 & 3 & 4 & 5 & 6\\
  \hline
Transient$+$pure & 1371 & 2884 & 1367 & 2256 & 1394 & 997 & 770\\
Pure & 356 & 1119 & 194 & 251 & 214 & 117 & 87\\
  \hline
  \end{tabular}
} \label{trod}
\end{table}

As an approximating function, we have chosen the power law

\begin{equation}
N = a q^b ,
\label{plaw}
\end{equation}

\noindent where $a$ and $b$ are two fitting parameters; $b < 0$.
We have also tried the exponential law $N \propto \exp (c q)$
(where $c < 0$), but it has turned out to be inappropriate, the
statistical significance of fitting being very low.

Since we are interested in the tail behaviour of the
distributions, we have used the data for $q \geq 1$, ignoring the
specific case $q=0$. In the case of all resonant
(transient-resonant plus pure-resonant) asteroids (see
Fig.~\ref{froa}) we find $a = 2821 \pm 454$, $b = -0.57 \pm 0.18$;
and the correlation coefficient $R^2 = 0.71$. As soon as $| b | <
1$, the predicted number of asteroids in high-order resonances is
formally infinite; in practice this means that it can comprise up
to $\sim 100\%$ of the whole set.

In the case of pure-resonant asteroids (see Fig.~\ref{fproan}) we
find $a = 1095 \pm 103$, $b = -1.64 \pm 0.28$; $R^2 = 0.94$. As
soon as $| b | > 1$, the predicted total number of asteroids in
high-order resonances is finite. Using Eq.~(\ref{plaw}), the
number of objects with $q \geq 7$ in the studied set is estimated
to be equal to $\approx 516$, or $\approx 22.1\%$ of the
identified number (2338). Therefore, the predicted total number of
asteroids in pure three-body resonances is estimated as 2854,
constituting $1.1\%$ of the whole set (249567).

A note of caution is in order here. One has to admit that the fits
made in this Section are based on few points, and, therefore, any
statistical predictions, made with these formulas, are uncertain.
What is more, one cannot even expect to find smooth distributions
and/or strict correlations in this field of research, where
dynamics is strongly interrelated with physics: the abundances of
objects in resonant groups are regulated by various processes,
e.g., by collisions and the Yarkovsky effect. Interactions with
asteroidal families are important. Of course, higher order
resonances should be directly analyzed in the future.

Another complicating factor is that the subresonance overlap (and
hence, the degree of chaoticity) in the multiplets is expected to
decrease asymptotically at $q \gg 1$. The reason is that the
subresonance typical width scales with $q$ as $\sim e^q$ (where
$e$ is the eccentricity), whereas the subresonances separation
(determining the multiplet width) remains basically constant,
depending solely on the secular precession rates of the
pericentres; thus the overlap/interaction of subresonances in the
multiplets decreases asymptotically at $q \gg 1$
\citep[p.~268--269]{NM99}. When the ratio of the subresonances
separation to the width of the leading subresonance is much
greater than one, the separatrix chaotic layers are exponentially
thin with this ratio, see \citep{C79,S08,S11}. It is
straightforward to suppose that, if one fixes the asteroidal
eccentricity, three-body resonances with increasing order become
basically regular \citep{NM99}. Thus, if the resonances are
populated uniformly in the eccentricity, the power-law
extrapolation for the high-order transient-resonant populations
fails beyond some order. Only if this critical order is high
enough, the extrapolation-based estimates might be appropriate.

\section{Lyapunov exponents in three-body resonances}
\label{Lestat}

The AstDyS database provides information on the maximum Lyapunov
exponents for almost all asteroids contained in it. The provided
values are computed on time intervals of 2~mln~yr, i.e., on time
intervals 20 times greater than that we use in our identification
procedure. It is instructive to check how the data on Lyapunov
exponents correlate with the pure/transient division (adopted in
our study) of the resonant asteroids.

In the transient case the motion is expected to be chaotic, and in
the pure case to be essentially regular, because transitions from
libration to circulation and vice versa are inevitable in a
separatrix chaotic layer of resonance. However, the adopted
relatively short integration time may imply that many
``pure''-resonant asteroids become transient on longer timescales.

\begin{figure}[!ht]
\includegraphics[width=\textwidth]{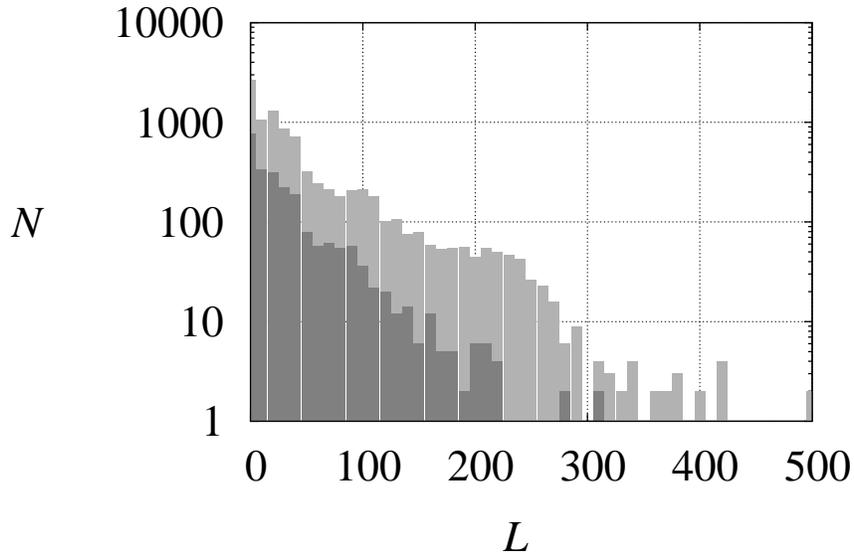}
  \caption{The distributions of resonant asteroids in the
  value of maximum Lyapunov exponent.
  Grey: asteroids in transient resonances.
  Black: asteroids in pure resonances.}
  \label{dist_LCE}
\end{figure}

First of all, let us build distributions of resonant asteroids in
$L$ (the maximum Lyapunov exponent). The constructed differential
distributions (histograms) are presented in a single plot in
Fig.~\ref{dist_LCE}. The Lyapunov exponents are given in units of
(mln~yr)$^{-1}$. (Note that in the AstDyS database they are given
in units of yr$^{-1}$.) $N$ is the number of objects in the
interval ($L$, $L + \Delta L$), where $\Delta L = 10$. The grey
histogram is for the asteroids in transient resonances, and the
black one is for the asteroids in pure resonances. The histograms
are cut off at $L=500$, because the objects with greater values of
$L$ are rare; they are all transients.

\begin{figure}[!ht]
\includegraphics[width=\textwidth]{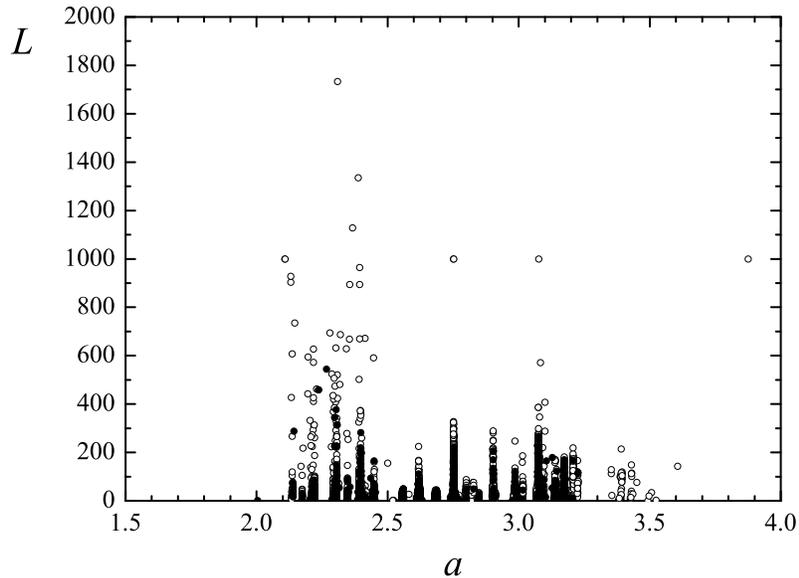}
  \caption{The maximum Lyapunov exponents versus the semimajor axis
  $a$.
  Open dots: asteroids in transient resonances.
  Black dots: asteroids in pure resonances.}
  \label{LCEs_a}
\end{figure}

From Fig.~\ref{dist_LCE} it is clear that the transients have a
much more extended distribution in $L$, in comparison with the
pure-resonant objects, as expected. This difference is uniform in
all resonant groups, as can be directly seen from
Fig.~\ref{LCEs_a}, where the maximum Lyapunov exponents are
plotted versus the semimajor axis $a$: the positions of open dots
(representing transients) extend to much greater heights, in
comparison with the pure-resonant objects, in all resonant groups.
Note that this plot is built without any cut-off in $L$, i.e., all
objects are shown. It is evident that only transients have very
large Lyapunov exponents, corresponding to Lyapunov times as small
as $\approx 570$~yr.

Among all identified pure-resonant objects, the average $L$ is
34.3~(mln~yr)$^{-1}$, and among all identified transients it is
49.7~(mln~yr)$^{-1}$. The Lyapunov times are $\approx 29200$ and
$\approx 20100$~yr, respectively. Among all objects of the AstDyS
database with measured Lyapunov exponents, the average $L =
20.8$~(mln~yr)$^{-1}$, corresponding to the Lyapunov time $\approx
48100$~yr. Thus both pure-resonant and transient-resonant objects
turn out to be more chaotic than a typical asteroid. The fact that
many pure-resonant asteroids have definitely non-zero Lyapunov
exponents signifies that they become transient on the timescales
longer than the adopted integration time ($10^5$~yr). This
underlines the conditional character of the ``pure/transient''
technical classification adopted in our study.

\section{Two-body resonances with Jupiter}
\label{sec_2br}

To form a more general statistical view of the actual resonant
structure of the main belt, it is instructive to compare the
abundances of asteroids in three-body resonances with the
abundances of asteroids in two-body resonances. For this purpose,
we have performed a procedure of identification of asteroids in
two-body resonances with Jupiter, analogous to that described
above for the case of three-body resonances. In this
identification procedure, we assume circular and zero inclination
orbits of perturbing planets.

Taking into account the D'Alembert rule, the resonant argument for
the resonance of order $q$ is defined  by the following formula
\citep{MD99, M02, Ga06}:

\begin{equation}
  \sigma = (p+q)\lambda_\mathrm{J}-p\lambda-q\varpi,
  \label{rad}
\end{equation}

\noindent where $\lambda_\mathrm{J}$ and $\lambda$ are the mean
longitudes of Jupiter and an asteroid, respectively, and $\varpi$
is the longitude of perihelion of the asteroid; $q$ is the
resonant order, $p$ is integer.

Of course, this setting of the identification problem is a rather
simplified one. When real resonant asteroids are considered, the
``circular'' subresonance term is not expected to dominate
universally; what is more, the resonant asteroid might reside in
another component of a multiplet (see \citealt{HM96,MH97}). We
regard the adopted approach as a first approximation, and leave
considerations of all possible subresonances for the future.

The resonant value of the semimajor axis of an asteroidal orbit is
given by

\begin{equation}
  a_\mathrm{res} \approx a_\mathrm{J}(1+\mu)^{-1/3}\left(\frac{p}{p+q}\right)^{2/3} ,
  \label{rsa}
\end{equation}

\noindent where $a_\mathrm{J}$ is the semimajor axis of Jupiter's
orbit, and $\mu$ is the mass of Jupiter in units of the mass of
the Sun (see, e.g., \citealt{MD99,Ga06}).

The identification matrix is constructed in the way analogous to
that described in Section~\ref{scim}. Similar to the case of
three-body resonances, it is assumed that $\gcd{(p, q)}=1$.
Besides, $1 \leq p \leq 11$.

The resonant order range is limited to $0 \leq q \leq 9$. Recall
that in the case of three-body resonances we have set $0 \leq q
\leq 6$. Choosing the upper bounds with the difference equal to 3
allows one to adequately compare identification statistics for
two-body and three-body resonances, because ``a three-body
resonance of a given order $q$ should have roughly the same
strength as a usual resonance of order $q+3$ for eccentricity of
about 0.05--0.10''~\citep{NM98}. Such a trick was used
in~\cite{NM98} for similar purposes.

\begin{figure}[ht!]
  \begin{center}
          \includegraphics[width=1.0\textwidth]{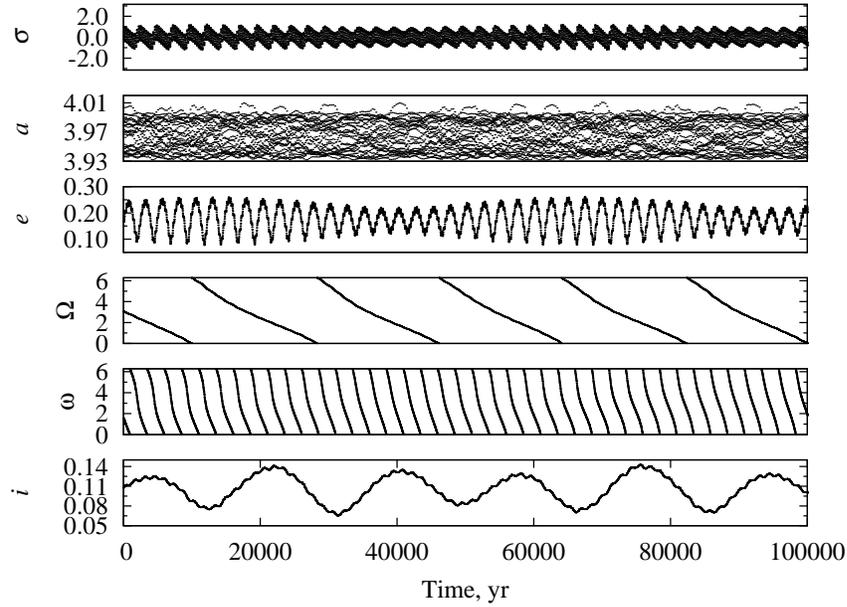}
    \caption{The orbital elements and resonant argument
  of 190 Ismene. Pure resonance 3/2.}
      \label{f190twb}
  \end{center}
\end{figure}

\begin{figure}[ht!]
  \begin{center}
          \includegraphics[width=1.0\textwidth]{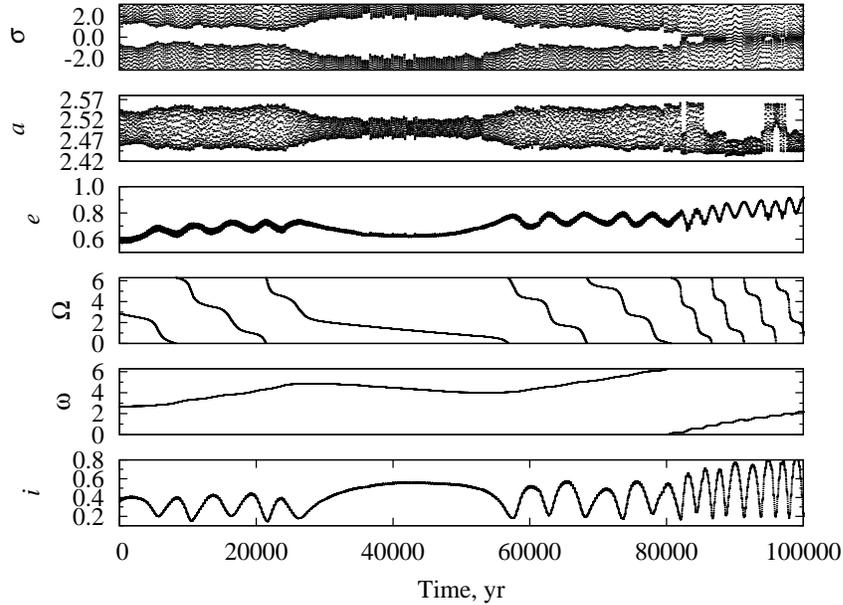}
    \caption{The orbital elements and resonant argument
     of 1915 Quetz\'alcoatl. Transient resonance 3/1.}
      \label{f1915twb}
  \end{center}
\end{figure}

Figs.~\ref{f190twb} and \ref{f1915twb} show the time behaviour of
the resonant argument and orbital elements of 1915 Quetz\'alcoatl
and 190 Ismene, as two typical examples. The top ten most
``populated'' resonances, identified in our study, are listed in
Table~\ref{tt10ptwbr}. All identified pure two-body-resonant
asteroids, grouped according to association to a given resonance,
are listed in the Appendix~B.\footnote{One may wonder why 279
Thule is absent in the 4/3 entry. The matter is that our automatic
procedure identifies it as being in transient resonance, not in
pure one, because the resonant argument sometimes goes out of the
range ($-\pi$, $+\pi$).}

The resulting statistics of identified objects are listed in
Tables~\ref{ttbrs} and \ref{tnrotb}. As follows from
Table~\ref{ttbrs}, a half of all identified asteroids in pure
two-body resonances are Trojans (1669/3132 $\approx 53\%$). Pure
Trojans plus pure Hildas constitute $\approx 85\%$ of all
asteroids in pure two-body resonances in our set. In
Table~\ref{tnrotb}, the $q$ dependence of the resonant asteroid
abundances is presented. The dependence is obviously irregular and
does not permit any smooth decay approximation. Especially one
should point out the negligible asteroid abundances at $q=3$ and
$q=9$.

\begin{table}[ht!]
\caption{The top ten most populated two-body resonances with
Jupiter. The main belt and Trojans}
\medskip \centering{
\begin{tabular}{rrrcrrl}
\hline
$p+q$ & $p$ & $q$ & $a_\mathrm{res}$ & Transient        & Pure & Notes \\
      &     &     & (AU)             &           + pure &      &       \\
\hline
 1 & 1 & 0 & 5.2043 & 1670 & 1669 & Trojan swarms  \\
 3 & 2 & 1 & 3.9716 & 1033 & 1007 & Hilda group    \\
 11 & 5 & 6 & 3.0766 & 351 & 49   &                \\
 2 & 1 & 1 & 3.2785 & 274 & 227   & Griqua family  \\
 8 & 3 & 5 & 2.7063 & 232 & 20    &                \\
 7 & 2 & 5 & 2.2576 & 208 & 20    &                \\
 10 & 3 & 7 & 2.3322 & 165 & 15   &                \\
 11 & 4 & 7 & 2.6514 & 165 & 0    &                \\
 9 & 4 & 5 & 3.0309 & 163 & 55    &                \\
 3 & 1 & 2 & 2.5020 & 63 & 32     & Alinda family  \\
 7 & 3 & 4 & 2.9583 & 51 & 17     &                \\
\hline
\end{tabular}
\label{tt10ptwbr}}
\end{table}

\begin{table}[ht!]
\caption{Asteroids in two-body resonances with Jupiter,
statistics. The main belt and Trojans}
\medskip \centering{
\begin{tabular}{lcc}
\hline
& Number               & Number            \\
&        of resonances &        of objects \\
\hline
Transient + pure & 21 & 4450 (1.78\%) \\
Pure & 16 & 3132 (1.25\%) \\
\hline
\end{tabular}
\label{ttbrs}}
\end{table}

\begin{table}[ht!]
\caption{Abundances of asteroids in two-body resonances in
function of $q$}
\medskip \centering{
\begin{tabular}{lrrrrrrrrrr}
\hline
 $q$ & 0 & 1 & 2 & 3 & 4 & 5 & 6 & 7 & 8 & 9\\
\hline
Transient + pure & 1670 & 1310 & 63 & 15 & 57 & 603 & 352 & 342 & 37 & 1\\
Pure & 1669 & 1236 & 32 & 5 & 19 & 95 & 49 & 23 & 4 & 0\\
\hline
\end{tabular}
\label{tnrotb}}
\end{table}

It turns out that in transient plus pure resonances the asteroids
are $\approx 2.5$ times more abundant in three-body resonances
than in two-body resonances (11039 versus 4450); and in pure
resonances the abundances are in the ratio $\approx 3:4$ (2338
versus 3132). However, if one excludes Trojans and Hildas, the
abundance of three-body-resonant asteroids becomes overwhelming:
in the case of transient plus pure resonances the ratio of
observed abundances of three-body-resonant objects and
two-body-resonant objects becomes equal to $11039/1747 \approx
6.3$; and in the case of pure resonances the ratio becomes equal
to $2338/456 \approx 5.1$.

If Trojans and Hildas are not excluded, the fraction of
pure-resonant asteroids among the transient+pure-resonant
asteroids is $\approx 3.3$ times greater in the two-body case than
in the three-body one ($\approx 70.4$\% versus $\approx 21.2$\%);
thus it might seem that the two-body resonances are much more
``pure'' on average. However, if one excludes Trojans and Hildas,
the fractions become comparable: $26.1$\% versus $21.2$\%.

One should outline that the comparative analysis given here is
merely statistical and, thus, formal. However, such an analysis
might provide a necessary preliminary stage for a deeper study of
the comparative role of two-body and three-body resonances; such a
study should comprise consideration of the physical (collisions
and the Yarkovsky effect) and dynamical (transport and diffusion)
processes, which lead to populating or depopulating the
resonances.

\section{Conclusions}

\begin{enumerate}

\item We have identified the resonant objects (the objects
residing in three-body resonances with Jupiter and Saturn in the
main asteroid belt) in the set of all numbered asteroids in the
AstDyS database. This set comprises 249567 asteroids catalogued up
to the date of April, 2011. The list of all asteroids identified
as residing in pure three-body resonances is given in Appendix~A.

\item The fraction of asteroids in three-body resonances
(transient plus pure) up to the 6th order inclusive turns out to
be $\approx 4.4\%$ of the total studied set of 249567 asteroids.
The fraction of asteroids in pure three-body resonances of the
same orders turns out to be $\approx 0.94\%$ of the total studied
set.

\item The top three most populated three-body resonances are: 5 -2
-2 (containing 699 transient+pure-resonant asteroids), 4 -2 -1
(688 transient+pure-resonant asteroids), 3 -2 -1 (621
transient+pure-resonant asteroids). For the pure-resonant
asteroids, the ``top three'' resonances are: 4 -2 -1, 3 -1 -1, and
5 -2 -2, containing 595, 203, and 182 objects, respectively.

\item Using a high-order extrapolation (in the form of a power
law) of the $q$ dependence of the number of identified resonant
objects, the actual total fraction of asteroids in pure three-body
resonances of all orders is estimated as $\approx 1.1\%$ of the
whole set. In what concerns the case of transient three-body
resonances, the situation is much less certain, because the
power-law extrapolation diverges.

\item We have also identified all objects residing in two-body
resonances (of order $0 \leq q \leq 9$) with Jupiter in the main
asteroid belt, taking the same database of asteroids. The list of
all asteroids identified as residing in pure two-body resonances
is given in Appendix~B.

\item The half of all identified asteroids in pure two-body
resonances are Trojans ($\approx 53\%$). The pure Trojans plus
pure Hildas constitute $\approx 85\%$ of all asteroids in the pure
two-body resonances. The $q$ dependence of the two-body resonant
abundances is clearly irregular and does not permit any smooth
decay approximation. Especially one should point out the
negligible asteroidal abundances at $q=3$ and $q=9$.

\item In the transient plus pure resonances, the identified
asteroids are $\approx 2.5$ times more abundant in the three-body
resonances than in the two-body resonances; and in the pure
resonances the abundances are comparable. However, if one excludes
Trojans and Hildas, the abundance of three-body-resonant asteroids
becomes overwhelming. What is more, taking into account
extrapolated abundances in high-order resonances may substantially
increase this overwhelming domination. Thus our analysis
quantitatively verifies the assertion by \cite{NM98} that ``the
three-body mean motion resonances seem to be the main actors
structuring the dynamics in the main asteroid belt''.

\item We would like to point out that our results confirm the
general concept of \cite{M68,M69} on the omnipresence
of resonances in the Solar system, however at a new level of
understanding of this phenomenon.

\end{enumerate}

\section*{Acknowledgements}

The authors are thankful to Zoran Kne\v{z}evi\'c and David
Nesvorn\'y for valuable remarks and comments.

This work was partially supported by the Russian Foundation for
Basic Research (project \# 10-02-00383) and by the Programmes of
Fundamental Research of the Russian Academy of Sciences
``Fundamental Problems in Nonlinear Dynamics'' and ``Fundamental
Problems of the Solar System Studies and Exploration''. The
computations were partially carried out at the St.\,Petersburg
Branch of the Joint Supercomputer Centre of the Russian Academy of
Sciences.

\newpage

\noindent {\large \bf Appendix A}
\bigskip

\noindent {\large \bf Asteroids in pure three-body resonances with Jupiter and Saturn}
\bigskip

\noindent
\textbf{1 3 -1 (3)}: 95 423 1102 3585 11456 17433 22499 27767 34493 41164 42441 46218 47896 48131 59909 62884 63667 72791 73569 74425 76056 78202 82703 91809 105839 112669 117065 117344 129360 138837 140653 141042 145390 155877 159321 161779 166668 186951 195633 195930 199688 200344 201223 201875 209429 209437 209473 210159 218079 226525 232332 232494 238387 239477 240226\newline\newline
\textbf{1 4 -1 (4)}: 9835 20596 31058 36595 37004 37611 41056 41640 50350 51002 55069 57488 58002 62632 66629 67442 71024 82600 83471 84938 87318 90368 93699 97281 107076 119544 122676 126319 130159 147292 158732 160021 166433 166516 167947 168639 178642 186262 188442 192877 212621 219368 226015 232815 235341 241349 249070\newline\newline
\textbf{1 6 -1 (6)}: 10482 12785 14833 16967 18960 25508 31374 33523 34780 39875 48030 53741 67456 74729 79395 82094 95170 105575 105721 130930 147022 180268 194048 199297 203374 210376 226238 230570 232392 233526 245785\newline\newline
\textbf{2 1 -1 (2)}: 1149 1670 38992 67811 105476 167454 229942 241160\newline\newline
\textbf{2 2 -1 (3)}: 70 194 258 839 923 995 16233 28706 37158 48104 48110 48127 48418 48532 51196 62082 67329 71407 84437 91337 93241 117209 122690 126001 133792 154671 173752 180242 186145 212288 224957 232199 238618 243865\newline\newline
\textbf{2 3 -1 (4)}: 7553 10368 39930 42942 48007 48028 48039 48105 48111 48118 48120 48121 48133 48509 48511 69468 70514 72106 74316 82341 89577 90015 90354 90546 112351 113728 115283 119372 122007 125642 129963 130692 150940 153684 157179 159616 161644 162752 167044 180486 180491 180493 180494 186531 188874 189427 197995 203427 216047 220119 222186 236300 239888 243110 244652 248126\newline\newline
\textbf{2 4 -1 (5)}: 9945 12169 15628 32302 32898 33946 38779 38993 39176 48014 48033 48103 48107 59305 75243 90851 99504 120401 138965 143630 168287 180218 211512 217139 218091 239263\newline\newline
\textbf{3 -2 -1 (0)}: 9864 10173 10926 17320 22785 26986 29849 29869 34592 36671 40940 41403 45312 47919 54497 56883 58847 58914 62989 64773 71551 73469 73570 73573 73798 73952 75981 76535 78756 82905 83011 83055 83103 83698 83972 84448 84640 90097 94202 97229 106316 112360 112384 112757 113085 113337 113402 113904 115445 123903 124012 127061 128255 128304 129408 130325 131350 132895 132912 133447 133714 135218 136290 138750 140372 140655 143617 144889 152112 157195 158147 160008 160150 164493 166768 166873 167567 169214 169298 170112 172023 173990 175484 182584 186114 189942 190125 195672 195779 195797 196016 196634 198642 199662 201338 201368 202406 203309 203815 204167 207764 209436 209446 209457 209458 209482 209487 209488 210167 210978 214804 216414 216642 217332 218433 221358 221408 223303 226565 227722 228301 231654 234751 234855 234957 235444 235958 240046 243227 243296 246676 247045 248205 249330\newline\newline
\textbf{3 -1 -1 (1)}: 564 947 1427 2042 2465 3534 4106 4426 5701 8068 9447 13085 13357 14169 14368 18302 19901 20118 20694 22146 22562 24011 24032 24190 25314 25891 27827 31192 33071 34104 35926 37025 41157 41868 42743 44362 44458 45486 48038 48539 50298 50954 54089 54405 54429 55893 56860 57161 57879 59296 59681 60160 61974 61980 62518 63423 64937 65543 67013 67028 67349 69300 69375 71281 71401 71423 71428 74215 74225 74314 76226 76238 76387 78768 79443 81841 81883 81927 83388 83935 87749 87893 88110 88173 89354 90647 93212 93749 95486 96994 97230 112181 114698 115643 117401 117668 118375 118835 120109 122501 122720 123213 123891 126305 127119 127499 131131 135585 138054 138096 139828 139925 140586 140605 140891 143738 146483 146507 147367 147370 148921 148949 148965 149535 152057 156621 157792 159133 161375 162113 164762 170012 170043 170655 171450 172327 172703 172908 174418 177037 179311 180571 180686 181134 181576 181761 181872 182146 185159 185934 188312 192832 195353 196624 198484 198567 198813 199291 200275 201564 201633 201847 202295 203717 204124 204800 206078 206622 206654 207269 207302 207661 208908 211245 212444 213406 217444 217785 217790 218768 218955 219401 220218 220322 222347 223648 223656 223845 225299 227701 229534 230119 231697 233422 234345 235689 239615 241729 242054 244969 245373 245400 247051\newline\newline
\textbf{3 1 -1 (3)}: 1705 6581 10851 15431 22249 36579 41431 44783 48124 50081 52125 52586 53090 58185 59601 64614 66163 68674 68996 69393 69636 74897 79142 80390 82322 88315 90730 97992 105977 114325 115219 118117 119371 125682 129748 129751 131646 134132 137237 139138 145918 153401 162608 165365 176110 177795 179798 180412 197692 205613 213218 213525 217478 220043 226224 232412 235665 240796\newline\newline
\textbf{3 3 -2 (4)}: 5413 10656 11653 21807 29946 41152 48515 48516 106548 106707 121883 138820 149695 182619 185797 209431 209440 209464 209469 209478 209483 209494 223389\newline\newline
\textbf{3 4 -2 (5)}: 28984 32932 33778 36169 39340 42104 56894 113850 116063 167477 168326 197436 209435 209439 209443 225316\newline\newline
\textbf{3 4 -1 (6)}: 209454\newline\newline
\textbf{4 -4 -1 (1)}: 52667 107282 129662 157240 163464 198658\newline\newline
\textbf{4 -3 -1 (0)}: 3870 7456 7530 19700 21151 22085 22686 22904 23678 23960 33719 35565 37541 37927 39547 39910 42708 48130 48535 50842 57528 61266 61284 61693 66867 68463 73206 75281 76150 76180 78509 81286 90130 90690 90721 93932 94135 96877 96966 99102 105019 105230 112405 113756 113933 119525 126653 126750 126821 132369 135793 138256 139761 144051 148896 149892 150968 158721 162842 165625 168566 169085 171254 172708 178376 179009 182622 182698 186902 187380 195541 196456 205418 210213 212226 214528 215264 221172 222747 227920 229825 236640 238272 240937 241763 241766 241991 242342 246382 248308\newline\newline
\textbf{4 -2 -1 (1)}: 463 2348 2487 2640 2791 3048 3293 3699 3716 3733 3865 4684 5156 5323 5747 6167 6193 6947 7133 7831 8044 8166 8772 9777 10009 10058 10092 10218 10371 10701 11011 11256 12065 12097 12161 12177 12299 12510 12674 12730 12985 13002 13024 13127 13159 13637 13711 14247 14823 15757 16163 16451 16525 16837 16867 17197 17363 17729 17869 18130 18319 18391 19029 19147 19619 20419 20658 20768 20775 20849 21556 22130 22150 22325 22326 22797 22829 22947 23907 24188 25715 26295 26665 26960 27260 27948 28451 29093 29442 30748 30799 30874 31466 31838 31873 31963 31971 32813 33804 34047 34707 36868 36943 38186 38695 39720 39958 39963 40379 40624 40635 40753 41005 41513 41697 42061 42081 42319 42626 42654 42666 43023 43499 44184 44224 45128 45256 45979 47257 48101 48513 48526 50597 51399 51412 51449 51563 51586 51628 52252 52374 52897 53095 53868 54082 54255 54357 54374 54698 54880 55658 56061 56612 56617 57668 57689 57772 57812 57822 58174 58601 59121 59413 59489 59721 60014 60217 60411 60790 60990 61119 61650 62098 63168 63620 64388 64567 64685 65540 65615 65800 66142 66165 66447 67019 67281 67938 68168 68183 68460 69470 70147 70224 70332 70346 70376 70409 70413 70572 71467 71977 72081 72596 73352 73540 73709 73802 73809 73875 74090 74589 74783 74981 75026 75067 75254 75317 75318 75362 75495 77034 77320 78298 78556 78942 79038 79126 79583 79713 80237 80658 80709 80798 81011 81017 81070 82585 84033 84098 84194 87288 88331 88580 88743 88817 89071 89512 89515 90121 90503 90515 90615 90977 90988 91319 92578 92605 92863 92879 93237 93238 94688 95119 95274 96372 96442 96683 98055 98186 98211 98416 99585 99777 105258 105378 105509 106090 106424 106862 106995 107111 112635 112731 113045 113416 113788 114358 114381 114393 114414 115067 115148 115739 116691 117187 118096 119087 119128 119129 119252 119322 119327 120754 122038 122121 122144 122171 122211 122987 124608 124893 124952 124961 125305 125404 125655 125661 126059 126235 126356 127513 127711 127725 128307 128375 129225 129343 129915 129940 130526 130622 130697 130774 130892 131004 131812 131974 132013 132106 132381 132566 133961 134582 134696 135643 136315 136959 137506 137576 137736 137873 137940 137946 138116 138134 139266 139282 140628 141811 141814 142007 142021 142132 142238 142240 142267 142288 142323 142338 142394 143253 145795 146052 146067 146075 146299 146300 146327 146986 147613 147643 147886 147993 148006 148501 149381 150148 150233 151518 152973 153023 153380 153752 154421 154749 154771 154903 155093 156074 156366 156875 156882 156905 158681 160447 160575 161210 161308 162728 162772 163114 164841 164849 165283 165371 165490 166087 166120 167081 167779 168144 168275 168802 168929 170267 170332 170346 171028 171534 171560 173167 173174 173265 173320 173620 174191 174824 175155 175400 175462 175491 175870 175885 175904 176707 177253 177465 178603 178843 179629 180430 180580 181100 181104 181801 182290 183153 183290 183786 183925 183934 183950 184459 185154 185710 186450 186857 187144 188383 189687 192429 192616 194342 194374 194425 194658 194663 194706 194804 194841 195074 195127 196258 196304 196390 196471 198228 198531 198942 200409 200893 200964 201028 201053 201117 202043 202257 202262 202504 202533 202949 203615 203902 204399 204470 205002 205218 205259 205852 205939 207042 207191 208256 208707 209615 210193 210554 210573 210761 211499 211558 211674 212440 213567 213608 213648 213882 214911 214927 215459 215461 215645 216082 216465 216617 217563 219123 219283 219317 219531 220210 220993 221782 221934 224069 224093 224121 224188 225672 226360 226694 228467 228477 228820 229366 229385 230586 230616 231235 231451 232038 232055 232810 234057 234532 234562 234583 235087 235098 237113 237509 237743 237831 238492 239586 239790 239898 240346 240581 240869 240898 241625 242784 243623 243785 243827 244214 244752 244778 246097 246158 246168 247378 247851 248673\newline\newline
\textbf{4 -1 -1 (2)}: 443 2175 2440 10364 20912 29435 30321 30916 53011 65455 67555 67699 76876 98342 98448 106572 116738 121837 124515 127569 129739 137412 141493 141594 143599 147582 149233 151938 171702 177786 178533 203286 210206 214109 214530 215602 217517 219155 224790 231602 242502\newline\newline
\textbf{4 1 -2 (3)}: 116793\newline\newline
\textbf{4 1 -1 (4)}: 5066 28441 141359 166908 209455\newline\newline
\textbf{4 3 -2 (5)}: 37957 62712 81077 81561 115442 126605 170321 180727 209153 213348 226790\newline\newline
\textbf{5 -7 -1 (3)}: 19282 22198 48500 58123 58525 59375 82464 82946 83299 84232 84696 87576 97536 105807 113343 118397 121291 130281 138591 141285 149124 155580 161209 161260 161771 168264 172761 179500 181007 189914 195767 195867 197461 201308 206843 208354 208477 209442 209444 209463 209471 209495 212905 217259 224768 237985 238056 242698\newline\newline
\textbf{5 -7 5 (3)}: 23539 26521 30985 71319 73466 76454 96142 106325 141402 145415 231035 240779\newline\newline
\textbf{5 -6 -1 (2)}: 5534 7176 14090 14941 17149 21960 29351 29871 30477 35153 35575 39299 41108 50657 53321 56310 62238 62810 68099 68723 69897 70592 70951 97743 105868 114380 121278 122957 123365 125842 132699 133224 139562 146555 147789 150481 153062 164540 170742 176979 177535 178321 195294 201559 221246 224675 230944 237117 243700 249153\newline\newline
\textbf{5 -6 5 (4)}: 43685 219393\newline\newline
\textbf{5 -4 -1 (0)}: 5277 9459 12814 13065 13757 16424 19419 22792 23520 23657 27100 27804 29650 30328 30679 30959 31439 32804 33412 35718 35991 36418 36424 38097 39810 42505 43479 44565 45181 45842 45888 46445 47258 48010 51708 51709 52841 52955 53791 55705 59106 59115 60633 61457 65886 66194 66269 66295 67385 67829 68668 69573 71753 71909 72327 75046 79865 80394 80850 88936 89454 92422 92524 97837 111991 112301 114347 115038 117889 118180 118353 121107 124291 124903 125165 125652 129793 131610 131808 137413 137578 137760 139104 139153 142094 145952 147019 153446 155114 155635 158205 160879 161918 163033 165230 166019 170983 171793 172567 172887 177239 179648 183044 183629 193928 194029 194089 194146 194297 201491 202894 205754 207455 208634 211549 212425 215432 216611 217155 217399 221620 222939 226260 229061 230577 232383 236196 238076 245780\newline\newline
\textbf{5 -3 -1 (1)}: 7882 29998 43935 106173 107454 207395 220455 221522\newline\newline
\textbf{5 -2 -2 (1)}: 490 744 755 786 818 1072 1209 1438 1701 1731 2039 2142 2164 2250 2492 2587 2666 2731 2757 2863 3460 4152 6830 9026 9163 11973 12762 12912 14360 16968 17165 17216 18202 21223 21547 21576 22097 25681 29519 29881 29931 31006 31748 32416 33225 33997 34070 35183 35660 38250 38507 40718 45587 46095 47652 47949 51926 52257 52693 53351 55682 57870 59160 62551 64977 66029 66990 68861 69977 70037 71628 73336 76793 79488 81891 81999 88003 90177 90263 90732 90749 94198 95638 97551 97733 97763 97772 99145 99254 106033 106113 114278 115826 121308 121556 121789 123740 124027 124040 127308 130258 132983 133482 134109 138716 141162 141215 141236 141351 143648 144814 149972 153256 153259 154145 155322 157670 159244 159526 159815 159972 168466 173044 173077 175803 177910 179478 181086 181160 183563 184199 189504 189885 195674 196076 196130 197452 198665 198785 201774 206892 209499 210746 214499 214566 214732 215565 217313 219134 219287 219935 221471 222187 223430 223461 223915 223926 223980 224643 224748 224749 227710 229258 229281 229285 230754 231394 232598 235502 236088 237605 239525 239696 240726 240783 242592 242713 242924 244065 248258 248482 248939\newline\newline
\textbf{5 -1 -2 (2)}: 952 26925 76782 84543 92035 94091 97702 105060 123639 129074 145233 163463 206169 209432 217709 218420 222105 222396 223884 229592 237971 242560\newline\newline
\textbf{5 -1 -1 (3)}: 97741 209474\newline\newline
\textbf{5 1 -2 (4)}: 30887 47629 48023 48117 59592 73345 75496 118814 126994 148536 149752 184749\newline\newline
\textbf{5 2 -2 (5)}: 3742 11237 48022 48138 48528 51225 75787 134216 147223 150902 166364 173403 180497 190240 192315 221121 227919\newline\newline
\textbf{5 3 -3 (5)}: 41188 51919 52365\newline\newline
\textbf{5 3 -2 (6)}: 28572 48012 48035 63365 69144 70259 77353 107672 118218 147173 153032 173280 177352 180405 180485 180487 180488 180490 180495 180496 180499 181592 195111 207078 208062 213634 217526 218932 220275 220992\newline\newline
\textbf{6 -7 -1 (2)}: 9905 36976 45278 48016 48137 53029 56026 61235 64335 65955 73093 94863 99656 128688 132181 134246 142870 145801 154396 158662 167079 180492 181579 183958 185852 189036 199811 202232 205870 207658 213096 214864 217403 224231 232466\newline\newline
\textbf{6 -7 5 (4)}: 21416 122051 180505 224163 228870\newline\newline
\textbf{5 4 -3 (6)}: 204141\newline\newline
\textbf{6 -6 -1 (1)}: 5684 60723 61210 77084 77140 78925 86835 92462 129754 130529 135577 145896 158281 161636 164255 168790 169644 177138 177673 192546 199074 216654 234556\newline\newline
\textbf{6 -4 -1 (1)}: 48001 97759 141744 157079 217839\newline\newline
\textbf{6 -3 -2 (1)}: 13797 19987 107393 122771 132844 204025 209370 237930 244227 246410\newline\newline
\textbf{6 -3 -1 (2)}: 76800\newline\newline
\textbf{6 -1 -2 (3)}: 3426 11792 20861 91347 98826 142623 148819 170636 196499 200520\newline\newline
\textbf{6 1 -3 (4)}: 57861 69553 76017 78776 116376 159801 168157 171677 184224 202322 209447 209449 209451 209453 209462 209480 209484 219975 232662 238386 240451\newline\newline
\textbf{6 1 -2 (5)}: 16601 23644 30162 41621 68649 89802 92860 99716 121093 132149 134702 139246 142066 150091 155432 163292 169468 188240 194745 197727 197853 211543 215096 220219 228253 230149 241162\newline\newline
\textbf{6 2 -3 (5)}: 4600 41401 51861 121579 159818 184675 209434 209467 244168\newline\newline
\textbf{7 -7 -2 (2)}: 15290 15731 23485 47732 113231 170857 177020 177095 181071 209450 212665 216400 236505 248433\newline\newline
\textbf{7 -7 -1 (1)}: 218044\newline\newline
\textbf{7 -6 -2 (1)}: 9838 29563 97767 106034 127629 140618 178715 196816 223929 248801\newline\newline
\textbf{7 -6 -1 (0)}: 33865 82020\newline\newline
\textbf{7 -5 -2 (0)}: 145459\newline\newline
\textbf{7 -4 -2 (1)}: 789 6473 9297 12276 30274 31230 51905 59439 62081 69966 71403 78848 89247 93712 94018 96004 97368 112284 117217 126886 132654 143342 150176 155000 178973 180648 181301 183437 184816 190375 207310 210646 213978 215744 215955 216219 218726 225563 228100 232806 236561 237914 238195\newline\newline
\textbf{7 -3 -2 (2)}: 16296 16431 37885 46792 47602 53099 61263 72777 88472 93131 117093 119590 148796 162187 178526 183397 202167 219327 225991 233905 239178\newline\newline
\textbf{7 -2 -3 (2)}: 1371\newline\newline
\textbf{7 -2 -2 (3)}: 10971 11148 16111 18757 20182 67775 70235 75374 95307 95778 126511 153543 156313 166240 177296 180507 183679 194470 222451 225756 228498\newline\newline
\textbf{7 -1 -2 (4)}: 20157 22519 27267 29032 33884 41473 48008 48027 48029 50177 51198 53766 57387 62411 72909 79157 79676 88742 95068 99518 125118 125567 130635 132085 132091 147946 154375 158303 159034 163148 170184 174793 175160 175831 213528 214159 218114 240121 245749\newline\newline
\textbf{7 1 -3 (5)}: 51664 57935 65732 116950 190207 223715\newline\newline
\textbf{7 1 -2 (6)}: 48020 48031 68793 131749 174692\newline\newline
\textbf{7 2 -3 (6)}: 63304 66880 79094 145551 219356 237227 237964 240175\newline\newline
\textbf{7 3 -4 (6)}: 120291\newline\newline
\textbf{8 -7 -2 (1)}: 7282 12592 31488 36675 69219 72509 77518 99201 123051 150021 157386 161417 166474 181601 182133 185382 186607 196547 204898 206018 219402 232499 236599 244538\newline\newline
\textbf{8 -4 -3 (1)}: 10 468 51250 56857 113931 134378 173978 209448 209485\newline\newline
\textbf{8 -3 -3 (2)}: 105866\newline\newline
\textbf{8 -3 -2 (3)}: 9791 21634 105959 106040 136430 181378 200353 225201\newline\newline
\textbf{8 -2 -3 (3)}: 151028 203738\newline\newline
\textbf{8 -1 -3 (4)}: 19698 41253 160914 185903\newline\newline
\textbf{8 -1 -2 (5)}: 181412\newline\newline
\textbf{8 1 -4 (5)}: 221458\newline\newline
\textbf{8 1 -3 (6)}: 4047 4553 7439 46087 96738 116854 125638 148867 152574 219335\newline\newline

\newpage

\noindent {\large \bf Appendix B}
\bigskip

\noindent {\large \bf Asteroids in pure two-body resonances with Jupiter}
\bigskip

\noindent
\textbf{1/1  (0)}: 588 617 624 659 884 911 1143 1172 1208 1404 1437 1583 1647 1749 1867 1869 1870 1871 1873 2146 2148 2207 2223 2357 2363 2456 2674 2759 2797 2893 2895 2920 3063 3240 3317 3391 3451 3540 3548 3564 3596 3708 3709 3793 3794 4007 4035 4057 4060 4063 4068 4086 4138 4489 4501 4707 4708 4709 4715 4722 4754 4791 4792 4805 4827 4828 4829 4832 4833 4834 4835 4836 4867 4902 5012 5023 5025 5027 5028 5041 5119 5120 5123 5126 5130 5144 5209 5233 5244 5254 5257 5258 5259 5264 5283 5284 5285 5436 5476 5511 5637 5638 5648 5652 5907 6002 6443 6545 6997 6998 7119 7152 7214 7352 7543 7641 7815 8060 8125 8241 8317 9023 9030 9142 9430 9431 9590 9694 9712 9713 9790 9799 9807 9817 9818 9828 9857 9907 10247 10664 10989 11089 11251 11252 11273 11275 11351 11395 11396 11397 11429 11487 11488 11509 11552 11554 11663 11668 11869 11887 12052 12054 12126 12238 12649 12658 12714 12916 12917 12921 12929 12972 12973 12974 13060 13062 13181 13182 13183 13184 13185 13229 13323 13353 13362 13366 13372 13379 13383 13385 13387 13402 13463 13475 13650 13694 13780 13782 13790 13862 14235 14268 14518 14690 14707 14791 14792 15033 15094 15398 15436 15440 15442 15502 15521 15527 15529 15535 15536 15651 15663 16070 16099 16152 16428 16667 16956 16974 17171 17172 17314 17351 17365 17414 17415 17416 17417 17418 17419 17420 17421 17423 17424 17442 17492 17874 18037 18046 18054 18058 18060 18062 18063 18071 18137 18228 18263 18268 18278 18281 18282 18940 18971 19018 19020 19725 19844 19913 20144 20424 20428 20716 20720 20729 20738 20739 20947 20952 20961 21271 21284 21370 21371 21372 21595 21599 21601 21602 21900 22008 22009 22010 22012 22014 22035 22041 22042 22052 22054 22055 22056 22059 22149 22180 22203 22227 22404 22503 22808 23075 23114 23118 23119 23126 23135 23144 23152 23269 23285 23355 23382 23383 23463 23480 23549 23622 23624 23694 23706 23709 23710 23947 23958 23963 23968 23970 23987 24018 24022 25344 25347 25883 25895 25910 25911 25937 25938 26057 26486 26510 26601 26705 28958 29196 29314 29603 29976 29977 30020 30102 30498 30499 30504 30505 30506 30510 30698 30704 30705 30708 30791 30792 30793 30806 30807 30942 31037 31342 31806 31819 31820 31821 31835 32339 32356 32370 32396 32397 32434 32435 32437 32440 32451 32461 32464 32467 32471 32475 32478 32480 32482 32496 32498 32499 32501 32513 32615 32720 32726 32794 32811 33822 34298 34553 34642 34684 34746 34785 34993 35272 35276 35277 35363 35672 35673 36259 36265 36267 36268 36269 36270 36271 36425 36624 36922 37297 37299 37300 37301 37519 37572 37685 37710 37714 37715 37716 37732 37789 37790 38050 38051 38052 38257 38574 38585 38592 38594 38596 38597 38598 38599 38600 38606 38607 38609 38610 38611 38614 38615 38617 38619 38621 39229 39264 39270 39275 39278 39280 39284 39285 39286 39288 39289 39292 39293 39369 39463 39474 39691 39692 39693 39793 39794 39795 39797 39798 40237 40262 41268 41340 41350 41353 41355 41359 41379 41417 41426 41427 42036 42114 42146 42168 42176 42179 42182 42187 42201 42230 42277 42367 42554 42555 43212 43436 43627 43706 45822 46676 47955 47962 47963 47964 47967 47969 48438 51339 51340 51344 51345 51346 51347 51348 51350 51351 51354 51357 51359 51360 51362 51364 51365 51378 51910 51935 51958 51962 51969 51984 51994 52273 52275 52278 52511 52567 52645 53418 53419 53436 53449 53469 53477 54581 54582 54614 54626 54632 54634 54638 54643 54645 54646 54649 54652 54653 54655 54656 54672 54678 54680 54689 55060 55267 55419 55441 55457 55460 55474 55496 55563 55568 55571 55574 55676 55678 55701 55702 56355 56951 56962 56968 56976 57013 57626 57714 57904 57910 57915 57920 58008 58084 58096 58153 58366 58473 58475 58478 58479 58480 58931 59049 59355 60313 60322 60328 60383 60399 60401 60421 61610 61896 62114 62201 62426 62692 62714 63175 63193 63195 63202 63210 63234 63239 63241 63257 63259 63265 63269 63272 63273 63278 63279 63284 63286 63287 63290 63291 63292 63294 64270 64326 65000 65097 65109 65111 65134 65150 65174 65179 65194 65205 65206 65209 65210 65211 65223 65224 65225 65227 65232 65240 65243 65245 65250 65257 65281 65583 65590 65811 67548 68112 68444 68725 68766 68788 69437 73637 73641 73677 73795 76804 76809 76812 76819 76820 76824 76826 76830 76834 76835 76836 76837 76838 76840 76857 76867 77891 77894 77897 77902 77914 77916 79444 80119 80302 80638 82055 83975 83978 83979 83980 83983 84709 88225 88227 88229 88240 88241 88245 88775 89829 89836 89841 89844 89852 89858 89871 89872 89878 89886 89898 89913 89918 89922 89924 89927 89934 89935 89938 89940 90337 90380 90569 96295 97893 97973 98037 98116 98139 98361 99306 99308 99309 99311 99323 99328 99368 99464 99943 99950 100475 100619 100624 101405 101466 101492 101612 103508 105720 105746 105803 105808 105896 105901 105904 106001 106060 106091 106160 107004 107134 107178 111932 114141 114208 114345 114694 114710 116134 116238 116439 116567 116901 116930 116954 116969 116970 117389 117395 117404 117423 117446 117447 117832 117851 118815 119528 120453 120454 122391 122460 122581 122592 122733 122860 122862 124695 124696 124729 124985 125045 125048 125059 125062 125106 125159 127532 127534 127846 128299 128301 128383 129130 129133 129134 129137 129140 129142 129144 129145 129147 129153 129207 129583 129602 130190 130588 130592 130687 131447 131460 131539 131546 131581 131635 133566 133853 133862 134251 134269 134329 134419 134720 134749 134957 134965 135540 135547 135593 135594 136557 137879 137954 138031 139009 141577 141584 150872 150876 151883 151884 152297 152517 152519 153083 153107 153122 153141 153155 153464 153500 153708 153755 153757 153758 154417 154632 154989 154990 154992 155326 155327 155332 155337 155415 155427 155786 155789 156098 156123 156125 156185 156188 156210 156237 156250 156252 156293 156294 156730 157468 157469 157470 157471 157740 157741 158231 158333 158336 158601 159162 159163 159340 159342 159378 159633 159658 160135 160140 160164 160465 160527 160528 160533 160534 160661 160679 160856 161003 161020 161024 161027 161044 161484 161489 161701 161717 162352 162380 162388 162396 162805 162811 162822 162851 162861 163102 163135 163155 163189 163196 163208 163216 163238 163256 163263 163731 164208 164210 164212 164585 165531 165551 165569 166115 166123 166136 166148 166211 166230 166285 166351 167686 168033 168280 168362 168364 171253 171424 171433 172819 173084 173086 173117 175129 175471 177655 178268 178291 178387 179133 179190 179217 179233 179244 179902 180265 180274 181278 181279 181648 181665 181751 181773 182121 182161 182163 182176 182178 182445 182487 182506 182516 182522 182548 182625 182627 182647 182666 182669 182675 182746 183309 183358 184274 184276 184280 184284 184306 184829 184937 184975 184976 184977 184978 184979 184980 184981 184982 184983 184984 184986 184988 185485 185486 185487 185489 185490 185492 185666 185693 185906 185919 186106 186125 186128 186138 186272 186544 187018 187019 187021 187024 187436 187456 187463 187469 187470 187471 187472 187473 187474 187475 187476 187478 187479 187607 187655 187657 187659 187692 187724 187755 188006 188020 188026 188053 188060 188234 188247 188257 188574 188835 188836 188837 188841 188842 188843 188844 188845 188846 188847 188942 188952 188965 188966 188976 188986 189004 189214 189215 189310 189377 189386 189493 189616 189772 189775 189859 190005 190229 190264 190267 190268 190294 190301 190309 190311 190351 190352 190353 190442 190446 192215 192217 192218 192220 192221 192222 192223 192224 192262 192268 192343 192345 192386 192388 192389 192390 192393 192423 192442 192448 192929 192938 192942 192964 193535 193570 193592 193602 193641 193670 195041 195084 195104 195117 195126 195152 195153 195167 195188 195217 195218 195230 195245 195258 195269 195273 195284 195286 195287 195308 195309 195312 195315 195318 195324 195337 195351 195490 195505 196316 196318 196364 196408 196440 196488 197563 197586 197593 197619 197620 197624 197628 197630 198858 199791 199792 199793 199796 200022 200023 200024 200027 200028 200029 200032 200035 200036 200037 200046 200051 200057 200544 201039 201040 202752 202756 202783 202791 202797 202824 202850 202854 202855 203210 203865 203891 204619 204847 204911 204927 205380 207749 210237 211992 212694 213180 213347 213351 213360 213394 214093 214376 214506 214508 214511 214514 215020 215107 215110 215199 215243 215319 215331 215340 215349 215407 215517 215530 215542 216034 216291 216292 216293 216307 216409 216419 216421 216423 216462 216847 216876 216877 216881 216883 216891 216898 217548 217612 217670 218070 218804 219057 219070 219139 219140 219141 219466 219834 219835 219837 219857 219866 219881 219890 219892 219896 219897 219902 219905 219907 219954 219958 220312 220318 220324 220333 220335 220336 220342 220351 220355 220356 220530 220555 220570 220574 220951 221542 221780 221785 221786 221908 221909 221910 221911 221912 221913 221914 221915 221917 221995 222049 222056 222063 222134 222379 222384 222773 222790 222791 222821 222826 222827 222831 222840 222843 222844 222847 222851 222854 222859 222861 222862 222864 222871 223238 223251 223268 223272 223615 223631 224020 224782 224792 224793 225089 225091 225209 225210 225211 225212 225213 225214 225216 225217 225218 225219 225220 225222 225223 225224 225226 225227 225230 225276 225294 225296 225318 225319 225359 225623 225624 225632 225984 225993 226005 226022 226027 226030 226033 226036 226053 226067 226348 226349 226352 226355 226358 226361 226367 226691 227759 228005 228006 228007 228008 228009 228010 228071 228077 228084 228085 228088 228089 228090 228095 228096 228097 228098 228099 228101 228102 228103 228104 228105 228106 228107 228108 228110 228111 228112 228113 228114 228115 228116 228117 228118 228119 228120 228121 228122 228123 228124 228125 228140 228150 228155 228162 228239 229760 229761 229808 229809 229813 229816 229822 229826 229832 229851 229859 229878 229902 229998 230339 231464 231493 231516 231572 231610 231612 231615 231620 231621 231623 231636 231664 231692 233578 233600 233683 233773 233790 233827 233915 233925 236878 236974 237035 237332 237338 237710 238623 239521 239903 240914 241099 241573 241576 241581 241582 241589 241643 241754 241956 241958 242177 242597 242624 242843 242917 243084 243088 243159 243313 243314 243315 243316 243318 243322 243323 243325 243334 243453 243460 243479 243484 243490 243494 243496 243501 243504 243509 245279 245682 246108 246113 246114 246145 246147 246523 246527 246530 246533 246534 246546 246550 246560 246817 246831 246833 246891 247019 247102 247297 247315 247323 247341 247351 247409 247421 247437 247653 247946 247967 247969 248048 248183 248186 248189 248238 248550 248575 248603 248670 248976 248978 248979 248985 248986 248990 248991 248996 249005 249032 249180 249212 249225 249236 249237 249241 249247 249256 249258 249465 249476 249481 249486\newline\newline
\textbf{2/1  (1)}: 1362 1921 1922 3688 3789 4177 5370 5528 8373 11097 11266 11573 11665 13963 14871 16882 18888 22740 23577 26553 28368 28459 29524 31249 31293 34901 35989 36140 37357 37528 37991 38406 38536 38584 38984 39018 39309 41262 42202 45511 45796 46168 46204 47547 51070 51267 51537 52700 55192 55222 55522 56326 57838 62679 62760 65297 65541 66174 68706 68738 71661 71694 73396 73972 73995 76694 77869 77911 78154 78801 78814 79282 79482 79500 82009 83904 83943 87362 87375 89175 89908 91182 91875 95942 97406 97653 98988 99686 100100 102419 102915 104742 107690 113723 114804 115916 116236 116559 117645 120907 120976 121675 121812 121824 124086 126119 127209 128313 128442 130295 130380 130385 131443 135454 135496 136807 138699 138739 138816 138854 140348 140956 141109 141127 141370 141394 143411 144443 145545 146258 146951 148125 149018 149613 152835 154621 154835 154904 154906 155030 155121 155874 155890 158152 159367 160013 161218 163660 165913 166564 166783 166912 167683 168337 169281 170869 171204 171414 172058 174006 174016 175976 178104 181067 182888 182931 185140 185483 185809 186358 186982 187001 187648 188795 188945 189375 190213 192606 193810 194142 196081 196114 196195 197406 197441 198770 198776 199440 199776 201364 202408 202771 203856 204265 205487 206286 206297 206298 208517 209024 209450 209489 209496 209499 214113 214461 217597 223012 225365 226587 228994 230800 232695 232726 232745 233635 234059 235455 235546 235547 236934 237665 243192 244606 245143 245382 246128\newline\newline
\textbf{3/2  (1)}: 153 190 361 499 748 958 1038 1162 1180 1202 1212 1268 1269 1345 1439 1512 1529 1578 1746 1748 1754 1877 1902 1911 1941 2067 2312 2483 2624 2760 2959 3134 3202 3254 3290 3415 3514 3557 3561 3571 3577 3655 3694 3843 3923 3990 4230 4255 4317 4446 4495 4757 5368 5439 5603 5661 5711 5928 6124 6237 6984 7027 7174 7284 8086 8130 8376 8550 8551 8743 8913 8915 9661 9829 10063 10296 10331 10608 10610 10632 11175 11249 11274 11388 11410 11542 11739 11951 12006 12307 12896 12920 13035 13317 13381 13504 13897 14195 14569 14669 14845 15068 15231 15278 15373 15376 15426 15505 15540 15545 15615 15638 15671 15783 16232 16843 16915 16927 16970 17212 17305 17428 17867 18036 19034 19752 20038 20628 20630 20640 21047 21128 21804 21930 22058 22070 22647 22699 23174 23186 23301 23405 25800 25869 26761 26929 27561 28918 29433 29591 29944 29973 30435 30764 31020 31097 31284 31338 31817 32395 32455 32460 32724 33753 34919 35016 35630 36182 36274 36941 37155 37452 37578 37590 38046 38292 38470 38553 38579 38613 38684 38701 38709 38830 39266 39282 39294 39301 39382 39405 39415 39427 40227 40238 40246 41278 41283 41365 41419 41488 42167 42190 42237 43818 43940 44549 45739 45850 45862 46302 46629 47907 48529 51178 51284 51298 51349 51838 51874 51885 51888 51930 51983 52702 54514 54599 54628 54630 54631 54644 54657 55196 55439 55498 55505 56982 56985 56996 57027 57759 58095 58188 58279 58353 59050 59079 59112 60232 60318 60381 60398 61042 62145 62241 62244 62408 62489 62820 62959 63184 63249 63293 63488 63491 64390 64739 64823 65236 65244 65374 65389 65821 65859 65989 66187 66227 67203 67246 67340 67368 68247 68287 68374 68402 68933 69302 69417 69566 73418 73436 73455 73457 73458 73475 73654 73769 73886 74051 74054 76750 76805 76810 76811 76822 76831 77734 77820 77884 77892 77893 77895 77903 77905 77910 78133 78159 78470 78477 78809 78815 79096 79097 79190 79439 79515 79724 82011 82023 82041 82043 82044 83722 83801 83804 83867 83877 83903 83916 84011 84103 87704 87811 87956 88230 88237 88246 88253 89928 90456 90502 90737 91273 91304 92272 92281 92283 92284 92287 92326 92344 94266 94275 94299 95952 95957 96086 96180 96609 97700 98002 99251 99276 99850 99862 99877 99982 100133 100229 100231 101153 101965 103537 104102 104943 105935 111928 111995 112432 112442 112553 112586 112694 112822 113224 113403 113415 114513 114749 114830 114837 114937 114954 114997 115088 115097 115170 115315 115380 115440 115463 116280 116294 116489 116512 116784 117108 117113 117131 117200 117288 117667 117993 118177 118520 119904 119918 119922 119935 119942 119944 119945 119946 119950 120175 120179 120287 120329 120336 120363 120618 120662 120761 120962 121005 121045 121074 124100 124174 124269 124791 125130 127426 127442 127448 127449 127469 127519 128059 128121 128126 128175 128198 128203 128209 128235 128254 128295 128420 128632 128858 129002 129007 129079 129241 129634 130453 130609 131421 131468 131481 131502 132868 132904 133324 133559 133614 134193 134233 134429 134562 134606 134642 134652 134690 134909 136822 136835 136935 137011 137014 137015 137312 137349 137816 138388 139799 141503 141518 141557 141701 142470 143621 143658 144851 144866 144894 145368 145373 145396 145397 145417 145421 145422 145426 145437 145443 145473 145478 145718 145767 145806 145833 145841 145843 145960 146014 146035 146961 147836 147846 147865 148060 148227 148234 149774 150033 150416 152094 152122 152132 152328 152373 152900 153357 153386 154262 154263 154264 154603 154629 154932 154960 154971 155125 155954 155959 156044 156085 156717 157014 157025 157076 157375 157710 157832 157861 157997 157999 158000 158158 158170 158194 158217 158595 158811 159339 159550 159937 160005 160234 160826 161334 162227 162232 162256 162280 162284 162314 162705 162777 162796 164843 164844 164858 164903 165044 165188 165306 165558 168517 168520 168525 168530 168545 168550 168557 169456 169508 169509 171258 171420 171465 171513 171562 171585 171600 172818 172965 172966 173012 173071 173083 173108 173351 173543 173643 174074 174077 174089 174114 174121 175115 175117 175350 175351 175355 175371 175372 175460 175620 175936 176158 176621 177640 177898 177940 177941 177943 178295 178299 181254 181256 181275 181618 181633 181647 181675 181778 181940 182062 182069 182094 182204 182239 183113 183579 184225 184241 184256 184780 184857 184930 184952 184955 184961 184963 185599 185640 185861 186448 186646 186649 187003 187011 187012 187017 187303 187378 187423 187477 187646 188188 188341 188342 188351 188830 188831 189114 189702 189714 189977 192203 193241 193248 193268 193291 193293 193306 193311 193323 193332 193350 193354 193398 193449 193501 194207 194384 194501 194512 194631 194724 194741 194742 194743 194860 194927 194989 195204 197478 197528 197537 197554 197558 198722 199715 200040 200043 200417 200441 200445 200465 200902 201937 202314 202551 202965 202975 202982 202992 203128 203156 203157 203159 204622 204803 204851 204856 204892 205070 205104 205134 207369 207379 207380 207381 207382 207384 207387 207605 207623 207644 207773 207827 207853 208015 208290 208695 209031 209507 209512 209919 209920 210049 210087 210149 210184 210187 210340 212337 212344 212345 212369 212490 212690 212928 214472 214483 214910 215089 215377 216408 216411 216613 216615 216771 217023 217024 217032 217039 217196 217208 217604 217844 217848 217849 217853 217877 217891 217906 217991 218241 219545 219555 220461 220910 221778 221982 222002 222427 222450 222454 222473 222488 222490 222517 222520 222521 222587 223211 223482 223492 223992 224008 224011 225765 225800 225868 225879 226233 226288 226295 226602 226612 227755 228093 228535 228559 228568 229050 229795 230251 230292 230293 230294 230309 230334 231400 231745 231749 231872 231909 231913 232088 232165 232393 232402 233778 233864 233878 233893 233939 233950 233980 234093 234650 234674 235567 235584 236100 236103 236114 236118 236119 236126 236209 236673 236677 236682 236819 236822 236852 236856 236861 236884 236894 236925 236973 237053 237253 237255 237278 237310 237321 237323 237849 237858 237861 238114 238154 238560 238804 238818 238819 238827 238829 238834 238847 239213 239341 239531 239692 239845 240134 241093 241376 241528 241563 241994 242008 242016 242148 242149 242159 242178 242290 242305 242309 242311 242321 242370 242451 242656 242682 242851 242996 243267 243311 243438 244397 244438 244478 244488 244878 245007 245093 245216 245578 246178 246496 246497 246578 246726 246786 247405 247507 247553 247563 247854 247879 247893 247900 247920 248004 248430 248945 249017 249060 249182 249246 249266 249275 249284 249302 249324 249391 249416\newline\newline
\textbf{3/1  (2)}: 887 6318 48012 48016 48017 48022 48024 48032 48034 48035 48040 48138 48140 86924 92664 131702 144861 152664 180485 180487 180488 180490 180492 180495 180496 180497 180498 180499 190119 207648 219527 220909\newline\newline
\textbf{4/3  (1)}: 185290 186024\newline\newline
\textbf{5/2  (3)}: 26760 26817 48038 214027 230979\newline\newline
\textbf{5/1  (4)}: 209454\newline\newline
\textbf{7/3  (4)}: 99169 141224 182274 209430 209432 209433 209434 209435 209439 209441 209443 209445 209452 209465 209466 209467 209470\newline\newline
\textbf{7/2  (5)}: 31010 48005 48006 48008 48010 48011 48014 48015 48020 48021 48026 48027 48029 48030 48031 48033 48036 48037 48126 203386\newline\newline
\textbf{8/3  (5)}: 9671 30843 32859 48019 48023 48139 104777 120251 125664 126492 129661 139227 150951 167506 168083 196568 210028 213384 226035 237892\newline\newline
\textbf{9/5  (4)}: 209490\newline\newline
\textbf{9/4  (5)}: 33016 47371 52714 74469 101642 105466 121614 128854 134821 150258 151436 163389 172417 173187 182682 186231 187691 187739 196684 200129 201196 209429 209436 209437 209438 209442 209444 209446 209447 209449 209456 209457 209458 209460 209463 209468 209471 209472 209473 209475 209476 209479 209482 209486 209487 209488 209491 209495 209497 230786 232988 237349 243711 245998 248481\newline\newline
\textbf{9/2  (7)}: 48001 175934 209455 209459 209474 209481 209492 209493\newline\newline
\textbf{10/3 (7)}: 48002 48004 48007 48009 48013 48018 48025 48028 48039 48115 180486 180489 180491 180493 180494\newline\newline
\textbf{11/5 (6)}: 4439 7992 10032 18105 26484 34952 38277 48003 57445 64959 99193 105840 106360 118403 120639 140934 146096 150539 162959 175111 176464 179413 186287 197393 198525 206171 208965 209431 209440 209448 209451 209453 209461 209462 209464 209477 209478 209480 209483 209484 209485 209494 209498 211894 213992 240485 242608 247708 248231\newline\newline
\textbf{11/3 (8)}: 65720 98009 141845 161599\newline\newline

\end{document}